\documentclass[journal]{IEEEtran}
\usepackage{graphicx}
\usepackage{latexsym}
\usepackage{amsfonts}
\usepackage{amssymb}
\usepackage{amsbsy}
\usepackage{color}
\usepackage{amsmath}
\usepackage{multicol}
\usepackage{float}
\usepackage{subfigure}
\usepackage{algorithm}
\usepackage{algorithmic}
\usepackage[dvips]{epsfig}

\usepackage{caption}
\usepackage{enumitem}
\usepackage{graphicx}
\usepackage{latexsym}
\usepackage{amsfonts}
\usepackage{amssymb}
\usepackage{amsbsy}
\usepackage{color}
\usepackage{amsmath}
\usepackage{multicol}
\usepackage{float}
\usepackage{subfigure}
\usepackage{algorithm}
\usepackage{algorithmic}
\usepackage[dvips]{epsfig}

\newcommand{\chapternote}[1]{%
 \let\thempfn\relax% Remove footnote number printing mechanism
  \footnotetext[0]{\emph{#1}}% Print footnote text
  }
\DeclareMathOperator*{\arginf}{\arg\inf}

\begin{document}

\title{Mean-Field Game Theoretic Edge Caching in Ultra-Dense Networks }

\author{\IEEEauthorblockN{\small Hyesung Kim, Jihong Park$^{\dagger}$, Mehdi Bennis$^{\dagger}$, Seong-Lyun Kim, and M\'erouane Debbah$^*$\\}
%\IEEEauthorblockA{\small
%School of Electrical and Electronic Engineering, Yonsei University,
%Seoul, Korea, email: \{hskim, slkim\}@ramo.yonsei.ac.kr\\
%$^*$Department of Electronic Systems, Aalborg University, Denmark, email: jihong@es.aau.dk\\
%$^\dagger$Centre for Wireless Communications, University of Oulu, Finland, email: bennis@ee.oulu.fi\\$^{\circ}$
%Mathematical and Algorithmic Sciences Lab, Huawei France R\&D, Paris, France \\
%and Large Networks and Systems Group (LANEAS), CentraleSup\'elec, Gif-sur-Yvette, France, email: merouane.debbah@huawei.com
%}

\thanks{H. Kim and S.-L. Kim are with the Radio Resource Management $\&$ Optimization
Laboratory, Department of Electrical and Electronic Engineering,
Yonsei University, Seoul, Korea (email: \{hskim, slkim\}@ramo.yonsei.ac.kr).}
\thanks{$^{\dagger}$J. Park and M. Bennis are with Centre for Wireless Communications, University of Oulu, 4500 Oulu, Finland (email: \{jihong.park, bennis\}@ee.oulu.fi).}
\thanks{$^*$M. Debbah is with Mathematical and Algorithmic Sciences Lab, Huawei France R\&D, Paris, France 
and Large Networks and Systems Group (LANEAS), CentraleSup\'elec, Gif-sur-Yvette, France (email: merouane.debbah@huawei.com).}
\thanks{This paper  was presented in part at IEEE International Conference on Communications (ICC) 2017 \cite{ICC}.}}

\maketitle

\begin{abstract}
This paper investigates a cellular edge caching problem under a very large number of small base stations (SBSs) and users. In this ultra-dense edge caching network (UDCN), conventional caching algorithms are inapplicable as their complexity increases with the number of small base stations (SBSs). Furthermore, the performance of UDCN is highly sensitive to the dynamics of user demand and inter-SBS interference. To overcome such difficulties, we propose a distributed caching algorithm under a stochastic geometric network model, as well as a spatio-temporal user demand model that characterizes the content popularity dynamics. By exploiting mean-field game (MFG) theory, the complexity of the proposed UDCN caching algorithm becomes independent of the number of SBSs. Numerical evaluations validate that the proposed caching algorithm reduces not only the long run average cost of the network but also the redundant cached data respectively by 24\% and 42\%, compared to a baseline caching algorithm. The simulation results also show that the proposed caching algorithm is robust to imperfect popularity information, while ensuring a low computational complexity.
\end{abstract}


%\linespread{2}

\section{Introduction}

5G cellular systems are envisaged to be extremely dense in order to cope with the relentless growth of user demand \cite{UDN_sur}--\cite{JensUdnMag}. A large number of small base stations (SBSs) in this ultra-dense network (UDN) should be supported by backhaul connections. The backhaul data traffic, increasing with the number of SBSs, may thus become the performance bottleneck. In this respect, edge caching is a promising solution, which alleviates the backhaul congestion \cite{caching_effect} by prefetching popular content at SBSs during off-peak hours \cite{Living}--\cite{caching_tradeoff:2016}, leading to an \emph{ultra-dense edge caching network (UDCN)}. 

Without edge caching, when a user requests a file, its serving SBS should first fetch  data from the file server via a wired backhaul link, and then wirelessly transmit  data to the user. On the other hand, with edge caching, the SBSs can directly transmit data to the users requesting the prefetched files, i.e. cache hitting users, thereby mitigating the backhaul congestion. In this paper, we investigate a UDCN, and seek an optimal content caching strategy that determines which target content files should be prefetched at each SBS equipped with a finite-sized caching storage. Due to the large number of SBSs, the caching strategy in UDCNs should accurately take into account the dynamics of the user demand and wireless network interference.

In fact, even a small misprediction of the user demand in a UDCN may result in a large number of the SBSs prefetching useless files. Conventional user demand models are thus too coarse to be applied, such as the Zipf's law model whereby the content popularity in the entire network follows a truncated power law \cite{Zipf_1}. Instead, the caching strategy design of a UDCN calls for a spatio-temporal user demand model that captures: long-term content popularity correlation and short-term popularity fluctuations, as well as the local popularity characteristics for different locations in the network. Furthermore, the downlink wireless interference induced by the SBSs in a UDCN may not only become large but also vary significantly for different network locations \cite{interf_udn}. As opposed to a traditional edge caching design relying on average interference under fixed SBS locations \cite{caching_avg_interf}, the UDCN caching strategy design thus necessitates an interference model reflecting spatial randomness of the caching SBSs and users. Unfortunately, it is obvious that incorporating these spatio-temporal dynamics of user demand and inter-SBS interference brings about a very complicated caching strategy. 

In order to reduce complexity, we propose a distributed UDCN caching strategy using a non-cooperative game theoretic approach, where SBSs minimize their own predefined cost function in a distributed fashion.
In our problem, each SBS determines its own caching decision so as to minimize the long run average (LRA) cost. The LRA cost of an SBS is an increasing function of (i) the wired backhaul utilization, (ii) the aggregate cached data size, and (iii) the amount of overlapping cached files among neighboring SBSs, i.e. content overlap, while decreasing with (iv) the wireless average downloading rate. The caching decision of each SBS is affected by other SBSs, through their cached files and inter-SBS interference, still incurring complication. Our UDCN caching strategy solves this  problem using two mathematical tools, mean-field game (MFG) theory and stochastic geometry (SG), respectively.

%\begin{figure*}
%\centering
%\includegraphics[angle=0, height=0.47\textwidth]{spatio_temp_ill}   %height=0.45
%\caption{\small{An illustration of the long-term and short-term temporal dynamics of user demand. The long-term dynamics are captured by the Chinese restaurant process, which determines the mean popularity for a certain time period $T$. During this period, the instantaneous popularity fluctuation is captured by the mean reversion model following the Ornstein-Uhlenbeck process  \cite{MF_ref2}.  }}\label{system_model_CRP} 
%\end{figure*}

To illustrate the effectiveness of our approach, we first consider the case without the aid of MFG and SG. Under the temporal content popularity dynamics, the optimal caching problem of a single SBS can be formulated as a stochastic differential game (SDG), which comprises a set of stochastic differential equations, known as Hamilton-Jacobi-Bellman equations (HJBs) \cite{exist_HJBsol1}. Solving these HJBs provides the optimal fraction of the files to be cached at the SBSs, thereby minimizing the LRA cost of each SBS. The key difficulty is that the number of these HJB equations scales with the number of SBSs. In addition, each SBS's HJB solution is intertwined with the HJB solutions of all the other SBSs, due to the aforementioned SBS interactions.

Instead of directly solving a large number of these coupled HJB equations, we consider the following approximation using MFG \cite{MFG_application}. Now, the interactions of each SBS with all other SBSs are approximated as the SBS's interaction with a single virtual agent acting like the aggregate SBSs. The interacting behavior of this virtual agent follows a stochastic distribution, known as mean-field (MF) distribution, which can be derived by solving a stochastic differential equation, called Fokker-Planck-Kolmogorov equation (FPK). As a result, we can find the optimal caching decision of each SBS from only a single HJB-FPK pair, regardless of the number of SBSs.

MFG guarantees that such an approximation always provides the optimal solution of the original problem, so long as the number of SBSs is very large while their aggregate interaction is bounded. A UDCN trivially satisfies the first condition. The latter can also be satisfied, since the inter-SBS interference of a downlink UDN is bounded by the SBS density, validated by using SG \cite{interf_var}. 

Next, we examine the average performance of the proposed UDCN caching strategy. With this end, we randomly select a user, i.e. a typical user, and calculate its LRA cost. Inter-SBS interference to this typical user can easily be analyzed using SG. Finally, the performance of the proposed UDCN caching algorithm can be calculated by solving a single HJB-FPK pair, capturing the interaction between a typical user's serving SBS and the virtual agent following the MF distribution. This simplification is visualized in Fig.~2, to be elaborated in detail in Section~III.

\subsection{Related work}

Edge caching has received significant attention in cellular networks \cite{caching_effect}--\cite{caching_tradeoff:2016}. Utilizing edge caches is  affected by user demand according to content popularity. The authors of  \cite{caching_dyna1}, \cite{MFG_caching} proposed a caching strategy that is adaptive to time-varying user request.
With regard to interference dynamics in caching networks, a recent study \cite{caching_dyna2} analyzed interference dynamics using SG for the spatial domain. In \cite{spatially_corr}, a spatially correlated content caching model was proposed for a given global content popularity information.
 However, there is no correlation between global and local popularity \cite{local_global}. This implies that the content popularity has spatial dynamics and that a caching strategy should be based on the local popularity information captured by each SBS.
On the basis of these preceding efforts, our work devises a caching control strategy that incorporates spatio-temporal dynamics.

When it comes to the UDN impact on edge caching, there are few previous investigations except \cite{MFG_caching} and our preceding work \cite{ICC}, which proposed a distributed caching algorithm considering temporal popularity dynamics in the UDN regime. 
It is worth mentioning that this study is rooted in  the UDN definition, where the number of SBSs exceeds
the number of users as in \cite{interf_udn}, \cite{JHP1}, \cite{JHP2}.
In UDN, both the interference distribution and the number of neighboring SBSs become different from those in traditional cellular networks. User's locations determine the interfering base stations inducing spatial dynamics of interference \cite{interf_udn}.                                                                                                                                                                                                                                                                                                                                                                                                                                                                                                                                                                                                                                                                                                                                                                                                                                                                                                                                                                                                                                                                                                                                                                                                                                                                                                                                                                                                                                                                                                                                                                                                                                                                                                                                                                                                                                                                                                                                                                                                                                                                                                                                                                                                                                                                                                                                                                                                                                                                                                                                                                                                                                                                                                                                                                                                                                                                                                                                                                                                                                                                                                                                                                                                                                                                                                                                                                                                                                                                                                                                                                                                                                                                                                                                                                                                                                                                                                                                                                                                                                                                                                                                                                                                                                                                                                                                                                                                                                                                                                                                                                                                                                                                                                                                                                                                                                                                                                                                                                                                                                                                                                                                                                                                                                                                                                                                                                                                                                                                                                                                                                                                                                                                                                                                                                                                                                                                                                                                                                                                                                                                                                                                                                                                                                                                                                                                                                                                                                                                                                                                                                                                                                                                                                                                                                                                                                                                                                                                                                                                                                                                                                                                                                                                                                                                                                                                                                                                                                                                                                                                                                                                                                                                                                                                                                                                                                                                                                                                                                                                                                                                                                                                                                                                                                                                                                                                                                                                                                                                                                                                                                                                                                                                                                                                                                                                                                                                                                                                                                                                                                                                                                                                                                                                                                                                                                                                                                                                                                                                                                                                                                                                                                                                                                                                                                                                                                                                                                                                                                                                                                                                                                                                                                                                                                                                                                                                                                                                                                                                                                                                                                                                                                                                                                                                                                                                                                                                                                                                                                                                                                                                                                                                                                                                                                                                                                                                                                                                                                                                                                                                                                                                                                                                                                                                                                                                                                                                                                                                                                                                                                                                                                                                                                                                                                                                                                                                                                                                                                                                                                                                                                                                                                                                                                                                                                                                                                                                                                                                                                                                                                                                                                                                                                                                                                                                                                                                                                                                                                                                                                                                                                                                                                                                                                                                                                                                                                                                                                                                                                                                                                                                                                                                                                                                                                                                                                                                                                                                                                                                                                                                                                                                                                                                                                                                                                                                                                          
The authors of \cite{MFG_caching} proposed a way to optimize caching control strategy  considering inter-BS interference in an UDN environment. 
It is worth noting that caching strategies of each SBSs are intertwined with all the above spatio-temporal dynamics of content popularity and network (interference and SBSs' backhaul and storage capacities). Therefore, a tractable method to evaluate the impact of the dynamics on a caching strategy needs to be investigated. Hence, the caching strategy can be jointly optimized subject to spatio-temporal dynamics and implemented in a decentralized~way.

%
%
%Note that our preliminary research about UDCN \cite{ICC} in part includes similar arguments to this article but under
%different scenarios. Specifically,  we considered the spatio-temporal dynamics (content popularity, interference, SBSs' backhaul and storage capacities)  and proposed a distributed caching algorithm that does not require the knowledge of other SBSs' caching control or states in \cite{ICC}. 
%Hence, we could achieve the low computational complexity regardless of the number of SBSs by exploiting MFG theory.

%This paper provides two major extensions (about temporal dynamics of user demand and computational complexity under multi-content caching scenarios) over our previous work \cite{ICC}, wherein we assumed that evolution law of time-varying popularity is independent of sequential user request. To make this model more practical, we suggest the temporal evolution process that incorporates the following sequential user behavior.
%Users sequentially request to download popular contents and the hits are recorded as history. In turn, this action influences the probability that the contents will be requested by other users.
%     
 
% None of the works on caching jointly considers the spatio-temproal dynamics of user demand and network (interference and SBSs' backhaul and storage capacities) in ultra-dense scenarios. 

%\pagebreak

\subsection{Contributions}

The main contributions of this paper are listed as follows
\begin{itemize} 

\item We propose a spatio-temporal popularity dynamics model (see Fig. \ref{system_model_CRP}), which captures the locally different content popularity but also the temporal content popularity changing within a long-term and short-term duration. 

\item A novel caching control algorithm for UDCNs is proposed, which minimizes the LRA cost while incorporating spatio-temporal demand and interference dynamics. Our algorithm is of low-complexity and independent of the numbers of SBSs, content files and users (see Proposition 1, and Fig. \ref{diagram}). We demonstrate the consistency of the algorithm for a large number of SBSs through simulation results (see Fig. \ref{complexity_it}). The proposed algorithm estimates other SBSs' caching strategies by predicting other SBS's states instead of directly gathering this information.

\item Numerical evaluation and analysis verify that the proposed algorithm achieves the unique mean-field equilibrium (MFE) \cite{MFG_application} (see Fig. \ref{MF_dist_total} and Proposition 1).

\item 
At the equilibrium state, numerical results demonstrate that the proposed algorithm reduces the LRA cost by almost 24\% compared to a baseline caching algorithm that does not consider the content overlap among neighboring SBSs but takes account of instantaneous content popularity, wired backhaul, storage, and interference. This performance enhancement grows when the user density increases (see Fig. \ref{LRA_user}).
Our algorithm also decreases the amount of overlapping contents per storage usage by 42\% (see Fig \ref{Repli}).

\item 
We also verify that the proposed MF caching algorithm is robust to imperfect popularity information through simulation results. Our algorithm produces lower LRA cost increment due to imperfect popularity information compared to the baseline caching algorithm without considering the amount of overlapping content (see Fig. \ref{LRA_incre}).
\end{itemize}

This article is structured as follows:
The network and dynamics models are described in Section II. Problem formulation and analytical results utilizing MF game theory and SG are also described in Section III.
The performance of the proposed algorithm is numerically evaluated in Section IV. Finally, 
concluding remarks are given in Section~V.

\section{System model and spatio-temporal dynamics}

\subsection{Ultra-dense network}

 We consider a UDN with SBS density $\lambda_b$ and user density $\lambda_u$. Their locations follow independent homogeneous Poisson point processes (PPPs), respectively. 
We set $\lambda_b \gg \lambda_u$, according to the definition of UDN \cite{interf_udn}.
User $i$ receives signals from SBSs within a reception ball $b(y_i,R)$ centered at the $i$-th user coordinates $y_i$, with radius $R$, which represents the average distance
that determines a region where the received signal power is larger than the noise floor, as depicted in Fig. 1. When $R$ goes to infinity, the reception ball model is identical to a conventional PPP network.  

Transmitted signals from SBSs experience path-loss attenuation. The attenuation from the $k$-th SBS coordinates $z_k$ to the $i$-th user coordinates $y_i$
is  $l_{k,i}=\min(1,||z_k-y_i||^{-\alpha})$, where $\alpha$ is the path-loss exponent.
The transmitted signals experience independent and identically distributed fading with the coefficient $g_{k,i}(t)$. We assume that the coefficient is not temporally correlated. The channel gain $h_{k,i}$ from SBS $k$ to user $i$ is  $|h_{k,i}(t)|^2=l_{ki}|g_{k,i}(t)|^2$. The received signal power is given as $S(t)=P|h_{k,i}(t)|^2$, where $P$ denotes the transmit power of an SBS.

The SBS directionally transmits signal by using $N_a$ number of antennas. Its beam pattern at a receiver follows a sectored uniform linear array model \cite{MIMO} where the main lobe gain is $N_a$ with beam width $\theta_{N_a} = 2\pi/\sqrt{N_a}$ assuming that side lobes are neglected and the beam center points at the receiver.

\subsection{Edge caching model}
\begin{figure}
\centering
\hspace{-20pt}\subfigure[Spatial popularity dynamics of the network at $t=0$.]{\includegraphics[width=8.5cm]{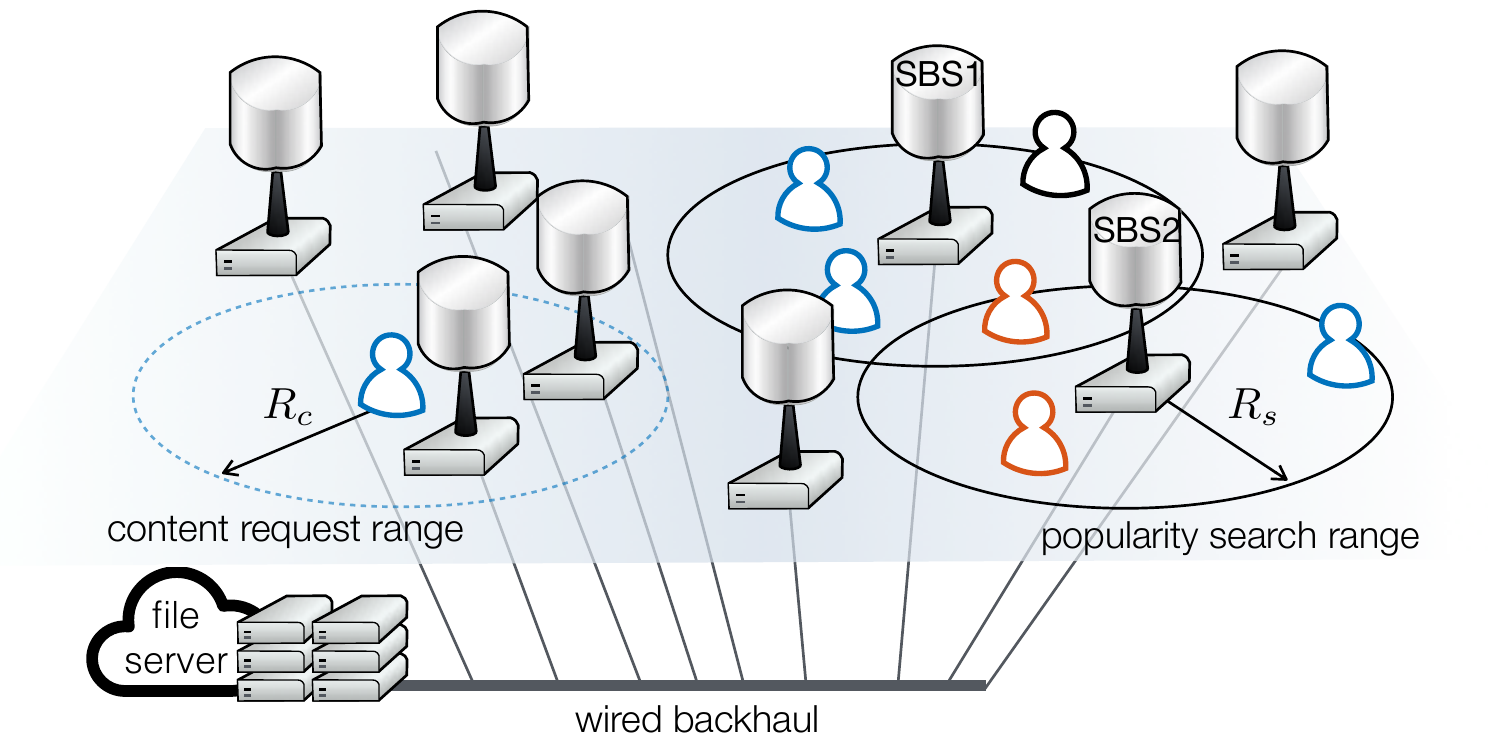}}\\
\subfigure[Temporal popularity dynamics at SBSs 1 and 2 during $3T$.]{\includegraphics[width=8cm]{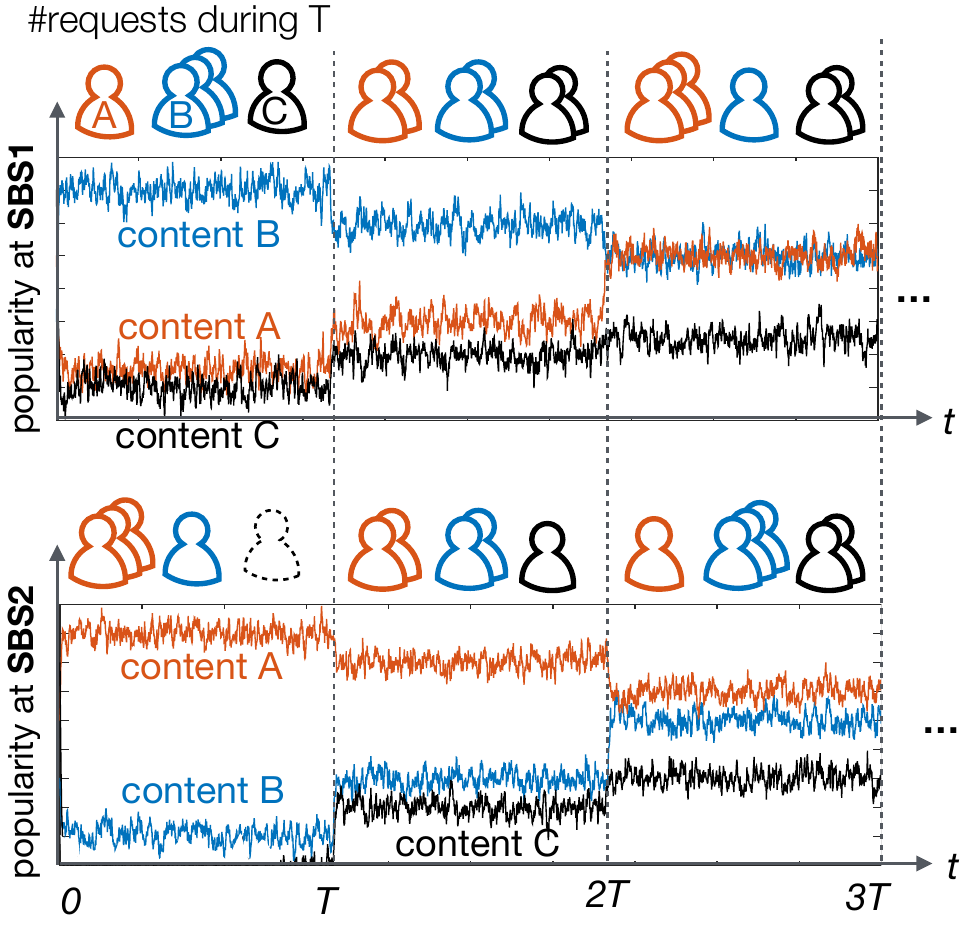}}
\caption{\small{An illustration of a UDCN and its spatio-temporal popularity dynamics. (a) Locally different popularity dynamics in  the UDCN.  (b) The content popularity changes within long-term and short-term duration. The long-term dynamics are captured by the Chinese restaurant process, which determines the mean popularity for a certain time period $T$. During this period, the instantaneous popularity fluctuation is captured by the mean reversion model following the Ornstein-Uhlenbeck process  \cite{MF_ref2}.  }}\label{system_model_CRP} 
\end{figure}

Let us assume that there is a set $\mathcal{N}$ consisting of $N$ SBSs within the reception ball with radius $R$.
Each SBS $k\in\mathcal{N}$ has data storage unit size $C_{k,j}$ assigned for  content $j$.
The storage units allow SBSs to download contents a priori from a server connected with non-ideal (capacity-limited) backhaul link  as depicted in Fig. \ref{system_model_CRP}. 
We assume that there is a set $\mathcal{M}$ consisting of $M$ contents, and the server has all the contents files in the set $\mathcal{M}$.
Users request  content $j$
from the contents set $\mathcal{M}$ with probability $x_j$, and the size of the content $j$ is denoted by $L_j$. 
The goal of SBS $k$ is to determine the amount of content files $\boldsymbol{p}_k(t)=\{p_{k,1}(t),...,p_{k,j}(t),...p_{k,M}(t)\}$, 
where $p_{k,j}(t) \in [0,1]$ is  a fraction of the file when SBS $k$ downloads content $j$  at time $t$. We assume that each content file is encoded using a maximum distance separable dateless code \cite{MDS}.

%The control variable $p_{k,j}(t)$ can also be interpreted as a fraction of the file in case that each content file is encoded using a maximum distance separable dateless code \cite{MDS}. 

According to the definition of UDN, multiple SBSs can serve a user \cite{interf_udn} as shown in Fig. 1.
We assume that a reference user is associated to one of the SBSs storing the requested content within a content request region as shown in Fig. \ref{system_model_CRP}a. If there are multiple SBSs having the same content within the region, the serving SBS is randomly chosen.
If no SBS has cached the requested content, one of the SBSs within the content request region fulfills the user request by downloading it from the server through the backhaul.
 %Fig. \ref{system_model} shows the system model.

%Consider small cell networks consisting of an SBS set $\mathcal{N}$ whose cardinality is $N$. 

%A reference user is served by an SBS with the cached content within a threshold distance $R_c$. 
%When there are multiple SBSs having the requested content within the region,
%we assume that one of them is randomly chosen to transmit the file to the user.
%This assumption allows us to make a set of SBSs that have same files, and that 
%set can be a player set for a game regime where strategies of SBSs are decoupled. 
% If no SBS has cached the requested content, one of the SBSs within $R_c$ fulfills the user request by download it from the server through the backhaul (caching service failed). 

 %This model is similar to caching model over SCNs in [\tred{Refer\_M}].

{\subsection{ Spatio-temporal dynamics of user demand on content and SBSs' states}}

In edge caching networks, caching strategies of each SBSs are intertwined with spatio-temporal dynamics of content popularity, SBSs' backhaul, storage capacities, and inter-SBS interference. In the following subection, we suggest tractable dynamics models to evaluate the impact of dynamics on the caching strategies.

\subsubsection{{\textbf{Spatio-temporal dynamics of user demand on content}}}

The user demand on a content varies over time and region according to its popularity. This spatio-temporal dynamics of the popularity can be illustrated in Fig. \ref{system_model_CRP}.

\textbf{Spatial dynamics}. In reality, user demand is different for each area, which means that there is no strong correlation between global and local popularity \cite{local_global}. Hence, a popularity dynamics model should capture the locally different content popularity.
To incorporate these spatial dynamics of popularity, we assume that each SBS has a popularity search region $R_s$ in which regional popularity
dynamics follows a stochastic process determined by sequential requests of users as shown in Fig. \ref{system_model_CRP}.  Therein, SBS 1 and SBS 2 have different sequential requests  and popularity dynamics for the same content A, B, and C. For tractability, let us assume that the stochastic processes are independent from each other. 
 The requests of subsequent users are influenced by the previous users' requests, in which users learn from users' previous selections. This sequential content request model is formalized in the followings.

\textbf{Temporal dynamics}. The temporal dynamics of popularity can be subdivided into long-term and short-term fluctuations \cite{MR_model}. 
The long-term dynamics of popularity determines a large popularity variation of each content. This observation gives rise to two contrasting phenomena that are illustrated in Fig. \ref{system_model_CRP}b. One is that popular contents have higher probability to be requested such as content A for SBS 2. The other is the rising in popularity of the up-to-date contents that have not been requested like content C for SBS 2. The short-term dynamics captures instantaneous small change of request probability and reflects a stochastic property for a short period of time.  
Specifically, a mean value of popularity changes periodically and maintains the mean value during a period in which the instantaneous popularity may fluctuate but will revert to its long-term mean value as shown in Fig. \ref{system_model_CRP}b.
 
For the long-term fluctuation, the stochastic process of subsequent content requests can be analogous to the table selection process known as Chinese Restaurant Process (CRP) \cite{Living}, \cite{CRP_1}. If we view the edge caching network as a Chinese restaurant, the contents as the tables, and the users as the customers, users sequentially request to download  popular contents and these hits are recorded as history. In turn, this recursive procedure makes popular contents requested more frequently. 

In this context, subsequent content request from users to each SBS can be modeled by an independent CRP where user request arrives randomly and independently at each SBS as shown in Fig. \ref{system_model_CRP}. We assume that SBSs get information of the long-term dynamics following CRP at every time period $T$ (i.e., $t=T, 2T,...,\kappa T,..., $ where $\kappa$ is a positive integer). 
Let us consider that there are two sets of contents for  SBS $k$. One is the set $U^r_k(\kappa T)$ consisting of contents that have been requested by $N_k(\kappa T)$ users at least once at SBS $k$ by time $\kappa T$. The other is the set $U^u_k(\kappa T)$ consisting of contents that have never been requested by time $\kappa T$. Let us define $n_{k,j}(\kappa T)$ as the number of requests for downloading content $j$ in the coverage of SBS $k$ by time $\kappa T$. For SBS $k$, the probability that a newly arriving user  decides to download contents is given according to the  following CRP \cite{CRP_1}:
\vskip 10pt
\begin{itemize} 
\item The user chooses a content that has never been requested with  probability  $\frac{\nu |U^r_k(\kappa T)|+\theta}{N_k(\kappa T)+\theta}$.
\item The user downloads a content $j$ previously requested according to the probability  $\frac{n_{k,j}(\kappa T)-\nu}{N_k(\kappa T)+\theta}$,
%, where $n_{k,j}(t)$ is the number of requests for downloading content $j$ in the coverage of SBS $k$ by time~ $t$.
\end{itemize}
where parameters $\theta$ and $\nu$ determine
correlation between the previous request history and a future request probability of a content. 
Each arriving user chooses a previously requested content file with probability proportional to the selection history $n_{k,j}(\kappa T)$, and downloads the unrequested content with a probability proportional to $\theta$ and $\nu$. 
The CRP model determines a mean popularity for a certain period where the content popularity varies with respect to the mean reversion model \cite{MR_model} as shown in Fig \ref{system_model_CRP}.  
The above CRP model determines a mean popularity value $\mu$ of a time period $T$, which is described as follows: 
\begin{equation}
\mu_{k,j}(\kappa T)  = \left\{ \begin{array}{ll}
 \frac{\nu |U^r_k(\kappa T)|+\theta}{N_k(\kappa T)+\theta} & \textrm{for $j \in U^u_k(\kappa T)$}\\
\frac{n_{k,j}(\kappa T)-\nu}{N_k(\kappa T)+\theta} & \textrm{for $j \in U^r_k(\kappa T)$}\end{array} \right.  \label{avg_Nk}
\end{equation}
For simplicity, we omit the index $\kappa T$ without loss of generality. Following \cite{MR_model}, we assume that the popularity varies within the period $T$ according to the mean reversion model using the Ornstein-Uhlenbeck (OU)  process \cite{MF_ref2} with a stochastic differential equation (SDE) as follows: 
\begin{equation}
\text{d}x_{k,j}(t)= r(\mu_{k,j}-x_{k,j}(t))\text{d}t+\eta \text{d}W_{k,j}(t), \label{SDE_x}
\end{equation}   
where $x_{k,j}(t)$ is the content request probability representing the popularity of content $j$ for SBS $k$ at time $t$ $(0 \leq t \leq T)$, $r$ is the change rate parameter, $\eta$ is a positive constant, and $W_{k,j}(t)$ is a Wiener process. Note that there exist different OU processes \eqref{SDE_x} depending on the corresponding static mean popularity $\mu_{k,j}$ for every time period T.
The SDE \eqref{SDE_x} represents a stochastic fluctuation of request probability around the mean value $\mu_{k,j}$. 
Due to the randomness of the term $W_{k,j}(t)$, the request probability $x_{k,j}(t)$ deviates from the long-term mean value $\mu$. If $x_{k,j}(t)$ becomes larger, the term $r(\mu_{k,j}-x_{k,j}(t))$ has a negative value and  makes $\text{d}x_{k,j}(t)$ head back to the mean value by complementing the random variation. 

%In the above long and short term dynamics models, if the long-term period $T$, CRP parameters $\theta$ and $\nu$ determine

%Every time the stochastic term $W_{k,j}(t)$ makes $x_j(t)$ deviate from the long-term mean value $\mu$, the deterministic term will act in such a way that $x_j(t)$ will head back to the mean value. 

%With that in mind, the behavior of the caching procedure is analogous to the table selection called Chinese Restaurant Process (CRP) \cite{Living}, \cite{CRP_1}. If we view the edge caching network as a Chinese restaurant, the contents as the tables, and the users as the customers, we can regard the contents request process by a CRP. In this model, users sequentially request to download  popular contents and the hits are recorded as history. In turn, this action influences the probability that the contents will be requested by other users. This recursive procedure makes popular contents requested more frequently. Meanwhile, there are definitely instances that non-popular contents are requested by users. 

\vskip 5pt
\subsubsection{\textbf{Temporal dynamics of cache storage size}} The remaining storage capacity varies according to the instantaneous caching strategy. 
Let us assume that SBSs have finite storage size and discard content files at a rate of $e_{k,j}$ from the storage unit in order to make space for caching other contents.
Considering the discarding rate, we model
the evolution law of the storage unit as follows: 
\begin{equation}
\text{d}Q_{k,j}(t)=(e_{k,j}-L_j p_{k,j}(t))\text{d}t, \label{SDE_q}
\end{equation}
where $Q_{k,j}(t)$ denotes the remaining storage size dedicated to content $j$ of SBS $k$ at time $t$, and $L_j$ is data size of content~$j$. Note that $L_jp_{k,j}(t)$ represents the data size of content $j$ downloaded by SBS $k$ at time $t$.

%To consider effect of wireless environment on caching strategy of each SBS, we use stochastic geometry to investigate average downlink spectral efficiency (SE), rate per unit bandwidth, under the following model.   
%
%The SBSs are uniformly distributed over a two-dimensional infinite plane with density $\lambda_b$, leading to a homogeneous Poisson point process (PPP). 
%Locations of users also follows a homogeneous PPP with density $\lambda_u$, but independent of SBS locations.
%An SBS having no serving user within its coverage does not transmit any signals, otherwise, it is active.
%
%User $i$ only receives signals from SBSs within a reception ball $b(y_i,R)$ centered at $y_i$ with radius $R$, which represents the average distance
%that determines a region where the received signal power is larger than noise floor. When $R$ goes to infinity, the reception ball model is identical to a conventional PPP network.  
%
%We assume that a reference user is served by a SBS with the cached content within a threshold distance $R_c$.
%When there are multiple SBSs having the same content $j$ within the region, the serving SBS is randomly chosen and transmit data to the user.  
%If no SBS has cached the requested content, one of the SBSs within $R_c$ fulfills the user request.
%Fig. \ref{system_model} shows the system model.

\vskip 5pt
\subsubsection{\textbf{Spatial interference dynamics}} In UDN, there is a considerable number of SBSs with no associated user within its coverage  which become idle and does not transmit any signal according to the definition of UDN ($\lambda_b\! \gg \!\lambda_u$) \cite{interf_udn}. Hence, this dormant SBS does not cause interference to neighbor SBSs. This leads to a spatially dynamic distribution of interference characterized by  users' locations.

We consider a randomly selected typical user and assume that active SBSs have always data to transmit. Let us denote the SBS active probability by $p_a$. The aggregate interference is imposed by the active SBSs with probability $p_a$. Assuming that $p_a$ is homogeneous over SBSs yields $p_a \approx 1-[1+\lambda_u/(3.5\lambda_b)]^{-3.5}$ \cite{SMYu}. It provides that the density of interfering SBSs is equal to $p_a\lambda_b$. Then, the aggregate interference of a randomly selected typical user is described as follows:
\begin{equation} 
\quad I^f(t)=\sum^{|\Phi_R(p_a\lambda_b)|}_k{ P|h_{k,i}(t)|^2},
\end{equation}
\noindent where $I^f(t)$ denotes the aggregate interference, and $\Phi_R(p_a\lambda_b)$ is a set of coordinates of active SBSs in the predefined  reception ball with radius $R$. 
Noting that $I^f(t)$ is a stochastic term, randomness of SBS locations induces the spatial dynamics of interference. 
Remark that the beam pattern at a receiver follows the sectored uniform linear array model where the main lobe gain is $N_a$ with beam width $\theta_{N_a} = 2\pi/\sqrt{N_a}$.
Then, the signal-to-interference-plus-noise (\textsf{SINR}) with $N_a$ number of transmit antennas is described as follows:
\begin{equation}
\mathsf{SINR}(t) = \frac{N_a P |h(t)|^2}{{\sigma^2+\frac{\theta_{N_a}}{2\pi}N_aI^f(t)}}. 
\end{equation}
\noindent  In the following section, we present the spatially averaged version of $I^f\!(t)$.

\section{Game theoretic problem formulation}

We utilize the framework of non-cooperative games  to devise a fully distributed algorithm. The goal of each SBS $k$ is to determine its own caching amount ${p}_{k,j}^*(t)$ for content $j$ in order to minimize an LRA cost. 
The LRA cost is determined by spatio-temporally varying content request probability,  network dynamics, content overlap, and aggregate inter-SBS interference. 
As the SBSs' states and content popularity evolves, the caching strategies of the SBSs must adapt accordingly. Minimizing the LRA cost under the spatio-temopral dynamics can be modeled as a dynamic stochastic differential game (SDG) \cite{MFG_application}. 
In the following subsection, we specify the impact of other SBSs' caching strategies and inter-SBS interference in the SDG by defining the LRA cost. 

%It leads to the backhaul link constraint as follows:
%\begin{equation}\label{constraint_bh}
%\sum_{j \in \mathcal{M}}B_{k,j}(t) \leq B_{k}(t),
%\end{equation}
 %Also, the amount of contents dowloaded by each SBS $k$ cannot exceed the available storage capacity $C_{k,j}$. 

\subsection{Cost Functions}

An instantaneous cost function $J_{k,j}(t)$ defines the LRA cost. It is affected by backhaul capacity, remaining storage size, average rate per unit bandwidth, and overlapping contents among SBSs. 
SBS $k$ cannot download more than $B_{k,j}(t)$, defined as the allocated backhaul capacity for downloading content $j$ at time $t$. 
We prevent the download rate $L_jp_{k,j}(t)$ exceeding the backhaul capacity constraint $B_{k,j}(t)$ by defining  the backhaul cost function $\phi_{k,j}$ as $\phi_{k,j}(p_{k,j}(t))\!=\! -\log (B_{k,j}(t)-L_jp_{k,j}(t))$. If $L_jp_{k,j}(t)\! \geq \!{B_{k,j}(t)}$, the value of the cost function $\phi_{k,j}$ goes to infinity, prohibiting the excess of the backhaul link. This form of cost function is widely used to model barrier or constraint of available resources as in \cite{MFG_caching}.
As cached content files occupy the storage, it causes processing latency \cite{Storage_cost} or delay to search requested files by users.    
This overhead cost is proportional to the cached data size in the storage unit. We propose a cost function for this storage cost with the occupation ratio of the storage unit normalized by the storage size as follows:
\begin{equation}
\psi_{k,j}(Q_{k,j}(t))= \gamma(C_{k,j}-Q_{k,j}(t))/{C_{k,j}},  \label{storage cost}
\end{equation}
where $Q_{k,j}(t)$ is storage cost function at time $t$.
\noindent Then, the global instantaneous cost is given by:
\begin{align}
J_{k,j}(p_{k,j}(t),\boldsymbol{p_{-k,j}}(t))\text{ }&=\text{ }\frac{\phi_{k,j}(p_{k,j}(t))(1\!+\!I^r_{k,j}(\boldsymbol{p}_{-k,j}(t)))}{\mathcal{R}_k(t,\hat{I}^f(t))x_j(t)}\nonumber\\
&+\text{ }\psi_{k,j}(Q_{k,j}(t)), \label{inst_global_cost}
\end{align}
\noindent where  $I^r_{k,j}(\boldsymbol{p}_{-k,j}(t))$ denotes the expected amount of overlapping content per unit storage size $C_{k,j}$, and 
$\hat{I}^f(t)$ denotes the normalized aggregate interference from other SBSs with respect to the SBS density and the number of antennas. The cost increases with the amount of overlapping contents and aggregate interference. The derivation of these two interactions is described in the next subsection. From the global cost function \eqref{inst_global_cost}, the LRA caching cost is given by:
\begin{equation}
\mathcal{J}_{k,j} = \mathbb{E} \left[\int_t^T J_{k,j}(p_{k,j}(t),\boldsymbol{p_{-k,j}}(t)) \text{ d}t \right]. \label{LRA cost}
\end{equation}
\subsection{Interactions: inter-SBS content overlap and interference}

The caching strategy of an SBS inherently makes an impact on the caching control strategies of other SBSs. We call these {\it interactions} among SBSs with respect to their own caching strategies. These interactions can be defined and quantified by the amount of overlapping contents and interference. These represents major bottlenecks for optimizing distributed caching for two reasons: first of all, they undergo changes with respect to the before-mentioned spatio-temporal dynamics, and it is hard to acquire the knowledge of other SBSs's caching strategies directly.
In this context, our purpose is to estimate these interactions in a distributed fashion without full knowledge of other SBSs' states or actions. 

\textbf{Content overlap.}
 As shown in Fig. \ref{system_model_CRP}a, in UDNs, there may be overlapping contents downloaded by multiple SBSs located within radius $R_c$ from the randomly selected typical user.
For example, let us consider that these neighboring SBSs cache the most popular contents with the intention of increasing the local caching gain (i.e., cache hit). Since only one of the SBS candidates is associated with the user to transmit the cached content file, caching the identical content of other SBSs becomes a waste of storage and backhaul usage.
In this context, overlapping contents increase redundant cost due to inefficient resource utilization \cite{social_cost1}. It is worth noticing that the amount of overlapping contents is determined by other SBSs' caching strategies. 
We define the content overlap function $I^r_{k,j}(\boldsymbol{p}_{-k,j}(t))$ as the expected amount of overlapping content per unit storage size $C_{k,j}$, which is given by:
\begin{equation}
I^r_{k,j}(\boldsymbol{p}_{-k,j}(t))=\frac{1}{C_{k,j}N_{r(j)}}\sum^{|\mathcal{N}|}_{i\neq k} {p}_{i,j}(t), \label{interaction_1}
\end{equation}
\noindent where $\boldsymbol{p}_{-k,j}(t)$ denotes a vector of caching control variable of all the other SBSs except SBS $k$, and $N_{r(j)}$ denotes the number of contents whose request probability is asymptotically equal to $x_j$. It can be defined as a cardinality of a set $\{ m | m \in {M} \text{ such that } |x_m-x_j| \leq \epsilon\}$. When the value of $\epsilon$ is sufficiently small, $N_{r(j)}$ becomes the number of contents whose request probability is equal to that of content $j$.
If there is a large number of contents with equal request probabilities, a given content is randomly selected and cached. Hence, the occurrence probability of content overlap decreases with higher diversity of content caching. 

%The diversity for content $j$ increases with the number of content $N_{r(j)}$ whose request probability is equal to content $j$. 

%Specifically, we suppose that $N_{r(j)}$ is the cardinality of a set $\{ m | m \in {M} \text{ such that } |x_m-x_j| \leq \epsilon\}$ where  $\epsilon$ is sufficiently small. 

\textbf{Inter-SBS interference.}
In UDNs, user location determines the  location of interferer, or the density of user determines the density of interfering SBSs. It is because there are SBSs that have no users in their own coverage and become dormant without imposing interference to their neighboring SBSs. 
These spatial dynamics of interference in UDN is a bottleneck for optimizing distributed caching such that an SBS in high interference environment cannot deliver the cached content to its own users. 
To incorporate this spatial interaction, following the interference analysis in UDNs \cite{JHP1}, interference normalized by SBS density and the number of antennas is given by: 
\begin{equation}
\hat{I}^f\!(t)\!=\!(\lambda_u\pi R)^2 N_a^{\!-\frac{1}{2}}\lambda_b^{-\frac{\alpha}{2}}\! \left(\!1\!+\! \frac{1-R^{2-\alpha}}{\alpha-2}\! \right)\!{P}\mathsf{E}_g [|g(t)|^2], \label{interaction_2}
\end{equation}
where $\hat{I}^f(t)$ denotes the normalized interference with respect to SBS density and the number of antennas.
%\begin{equation}
%{I}^f_{p_a\lambda_b}f(t)\!=\!(\lambda_u\pi R)^2 N_a^{\!-\frac{1}{2}}\lambda_b^{-\frac{\alpha}{2}}\! \left(1+ \frac{1-R^{2-\alpha}}{\alpha-2} \right){P}\mathsf{E}_g [|g(t)|^2]. \label{interaction_2}
%\end{equation}
It gives us an average downlink average rate per unit bandwidth $\mathcal{R}_k(t)$ and its upper bound in UDN as follows:
\begin{eqnarray}
\mathcal{R}_k(t)&=&\mathsf{E}_{S,I^f}\left[\log(1+\mathsf{SINR}(t))\right] \\ &\leq& \mathsf{E}_{S}\log\left(1+\frac{S_k(t)}{\frac{\sigma^2}{ N_a \lambda_b^{\alpha /2}} + \mathsf{E}_{I^f}[{\hat{I}}^f(t)]}\right), \label{ER_1}
\end{eqnarray}
\noindent where $\sigma^2$ is the noise power. Note that inequation \eqref{ER_1}  showes the effect of interference on the upper bound of an average SE. It is because we consider that only the SBSs within the pre-defined reception ball cause interference to a typical user. Hence, the equality in inequation \eqref{ER_1} holds, when the size of reception ball $R$ goes to infinity, including all the SBSs in the networks as interferers.

\subsection{Stochastic Differential Game for Edge Caching}

As the SBSs' states and content popularity evolves according to the dynamics \eqref{SDE_x} and \eqref{SDE_q}, an individual SBS's caching strategy must adapt accordingly. Hence, minimizing the LRA cost under the spatio-temporal dynamics can be modeled as a dynamic stochastic differential game (SDG), where the goal of each SBS $k$ is to determine its own caching amount ${p}_{k,j}^*(t)$ for content $j$ in order to minimize the LRA cost  $\mathcal{J}_{k,j}(t)$ \eqref{LRA cost}. 
\begin{equation}
\hspace{-70pt} \textbf{(P1)} \qquad\qquad\quad\quad v_{k,j}(t)=\mathop{\text{inf}}\limits_{p_{k,j}(t)}\text{ } \mathcal{J}_{k,j}(t).
\end{equation}
\vskip -13pt
\begin{flalign}
\text{subject to }\quad\quad&\text{d}x_j(t)= r(\mu-x_j(t))\text{d}t+\eta \text{d}W_j(t), \quad\label{const_1} \\ 
&\text{d}Q_{k,j}(t)=(e_{k,j}-L_j p_{k,j}(t))\text{d}t.   \label{const_2}
\end{flalign} 

In the problem \textbf{P1}, the state of SBS $k$ and content $j$ at time $t$ is defined as 
$\boldsymbol{s}_{k,j}(t)=\{x_j(t),\mathcal{R}_k(t),Q_{k,j}(t)\}$, $\forall k \in \mathcal{N}, \forall j \in \mathcal{M}$. The stochastic differential game (SDG) for edge caching is defined by
$(\mathcal{N},\mathcal{S}_{k,j}, \mathcal{A}_{k,j}, \mathcal{J}_{k,j}    )$ where $\mathcal{S}_{k,j}$ is the state space of SBS $k$ and content $j$, $\mathcal{A}_k$ is the set of all caching controls $\{p_{k,j}(t), 0 \leq t \leq T \}$ admissible for
the state dynamics.

In order to solve the problem \textbf{P1}, the long-term average of content request probability $\mu$ is necessary in the dynamics of content request probability \eqref{const_1}. 
To determine the value of $\mu$, the mean value $m_k(t)$ of the cardinality of the set $U_k^r(t)$ needs to be obtained. 
Although the period $\{0\leq t\leq T\}$ is not infinite, we assume that the inter-arrival time of content request is sufficiently smaller than $T$ and that numerous content requests arrive during that period. Then, the long-term average of content request probability $\mu$ becomes an asymptotic mean value $(t\!\rightarrow\! \infty)$.
Noting that $\sum_j n_{k,j}(t) = N_k(t)$, the mean value of $m_k(t)$ is aymptotically given by \cite{CRP_2} as follows:
\begin{equation}
\left<|U^r_k(t)|\right>  \simeq \left\{ \begin{array}{ll}
 \frac{\Gamma(\theta+1)}{\alpha\Gamma(\theta+\alpha)}N_k(t)^{\alpha} & \textrm{for $\alpha>0$}\\
\theta \log(N_k(t)+\theta) & \textrm{for $\alpha=0$}\end{array} \right.  \label{avg_Nk}
\end{equation}
where the expression $\left<.\right>$ is the average value, and $\Gamma(.)$ is the Gamma function.

%, and $\mathcal{J}_{k,j}$ is the LRA cost over a time window $[0,T]$ defined as follows:
%
%{
%\begin{equation}
%\mathcal{J}_{k,j} = \mathop{E} \left[\int_t^T J_{k,j}(p_{k,j}(t),\boldsymbol{p_{-k,j}}(t)) \text{ d}t + \kappa(\underline{\boldsymbol{s}}_{k}(T))\right],
%\end{equation}}
%
%\noindent where $\kappa: \mathcal{S}_{k,j} \rightarrow \mathbb{R}, \underline{\boldsymbol{s}}_{k} \mapsto \kappa(\underline{\boldsymbol{s}}_{k})$ is the cost of having a remaining storage size at the end of the time window $[0,T]$. 
%Thus, the SDG is formulated as follows:

%\begin{eqnarray} \label{eq:p1}
%&&\hspace{-40pt}\textbf{(P1)}\quad\mathop{{\text{minimize}}}\limits_{p_{k,j}(t)}\text{ } \mathcal{J}_{k,j}\\
%&&\hspace{-40pt}\quad\quad\quad\text{such that}\text{ }\text{ } \text{d}x_j(t)= (u_j-a_j)\text{d}t+\eta \text{d}W_j(t), \label{eq:p1c1}\\
%&&\hspace{-40pt}\text{ }\qquad\quad\quad\quad\qquad \text{d}Q_{k,j}(t)=(e_{k,j}-L_j p_{k,j}(t))\text{d}t. \label{eq:p1c2}
%\end{eqnarray}

%\begin{equation}
%\hspace{-70pt} \textbf{(P2)} \qquad\qquad\quad\quad v_{k,j}(t)=\mathop{\text{inf}}\limits_{p_{k,j}(t)}\text{ } \mathcal{J}_{k,j}(t).
%\end{equation}
%\begin{flalign}
%\text{subject to }\quad\quad&\text{d}x_j(t)= f(x_j)\text{d}t+\eta \text{d}W_j(t), \quad\label{const_1} \\ 
%&\text{d}Q_{k,j}(t)=\sum_j\left[e_{k,j}-L_j p_{k,j}(t)\right]\text{d}t.   \label{const_2}
%\end{flalign} 

The problem \textbf{P1} can be solved by using a backward induction method where the minimized LRA cost $v_{k,j}(t)$ is determined in advance through solving the following  $N$ coupled HJB equations. 
\begin{align}
0&\!= \partial_t v_{k,j}(t)\! +\!\mathop{\text{inf}}\limits_{p_{k,j}(t)} \bigg{[}J_{k,j}(p_{k,j}(t),\boldsymbol{p_{-k,j}}(t))\!+\! \frac{\eta^2}{2}\partial_{xx}^2 v_{k,j}(t)\nonumber \\
 &+\! \underbrace{(e_{k,j} -L_j p_{k,j}(t))}_{(A)} \partial_{Q_k}v_{k,j}(t)+\underbrace{r(\mu-x_j(t))}_{(B)}\partial_x v_{k,j}(t)\bigg{]}. \label{hjb_sdg}
\end{align}
If there exists a joint solution of the HJB equations \eqref{hjb_sdg},
the existence of a Nash equilibrium of the problem \textbf{P1} is guaranteed \cite{exist_HJBsol1}. 
Since the smoothness of the drift functions defining temporal dynamics (A) and (B) and the cost function \eqref{inst_global_cost} are respectively satisfied \cite{exist_HJBsol1}, the partial differential equations (PDEs) \eqref{hjb_sdg} have a unique joint solution $v^*_{k,j}(t)$ \cite{exist_HJBsol1}.
Thus, there exists a unique minimal cost $v^*_{k,j}(t)$ of the problem \textbf{P1} and its corresponding Nash equilibrium (NE) \cite{MFE_def} defined as follows:

\vskip 5pt
\noindent {\bf Definition 1}: The set of SBSs' caching strategies $\mathbf{p}^*=\{p_{1,j}^*(t),...,p_{N,j}^*(t) \}$, where $p_{k,j}^*(t) \in  \mathcal{A}_{k,j}$ for all $k \in \mathcal{N}$, is a Nash equilibrium, if for all SBS $k$ and for all adimissible caching strategy set $\{p_{1,j}(t),...,p_{N,j}(t) \} $, where $p_{k,j}(t) \in  \mathcal{A}_{k,j}$ for all $k \in \mathcal{N}$, it is satisfied that 
\begin{align}
\mathcal{J}_{k,j}(p_{k,j}^*(t), \boldsymbol{p}_{-k,j}^*(t))\leq \mathcal{J}_{k,j}(p_{k,j}(t), \boldsymbol{p}_{-k,j}^*(t)), \label{Def_NE}
\end{align}
under the temporal dynamics \eqref{const_1} and \eqref{const_2} for common initial states $x_j(0)$ and $Q_{k,j}(0)$.

\vskip 5pt

Unfortunately, this approach is computationally complex to be implemented in achieving the NE \eqref{Def_NE}, when $N$ is larger than two, because an individual SBS should take into account other SBSs' caching strategies $\boldsymbol{p_{-k,j}}(t)$ 
to solve the inter-weaved system of $N$ HJB equation \eqref{hjb_sdg}.  Furthermore, it requires collecting the control information of all other SBSs including their own states, which brings about a huge amount of information exchange among SBSs. This is not feasible and impractical for UDNs. For a sufficiently large number of SBSs, this problem can be transformed  to a mean-field game (MFG), which can achieves the $\epsilon$-Nash equilibrium \cite{MFG_application}.

%\begin{figure*}\centering
%\subfigure[A]{
%\includegraphics[angle=0, height=0.22\textwidth]{MFdist_x_03}}   %height=0.33
%\subfigure[B]{
%\includegraphics[angle=0, height=0.22\textwidth]{MFdist_x_06}}   %height=0.33
%\subfigure[C]{
%\includegraphics[angle=0, height=0.22\textwidth]{MFdist_x_09}}   %height=0.33
%\end{figure*}

% With utilizing MFG and SG, system does not need interfaces to get full knowledge of other SBSs' caching control. The number of equations to solve is reduced to two from $N$.
%\noindent
%\begin{figure*}
%\centering
%\includegraphics[angle=0, width=0.9\textwidth]{MF_dist_total2} 
%\caption{Evolution of the MF distribution $m^*_t(Q)$ of the remaining storage with respect to different content popularity $x$ $(B=1, \mu =0.1)$. The initial MF distribution $m_0$ is given as $\mathcal{N}(0.7,0.05^2)$. }\label{MF_dist_total}
%\end{figure*}

\subsection{Mean-field Game for Caching}

It is hard to react to each individual SBS's caching strategy when solving the SDG \textbf{P1}.
Fortunately, when the number of SBSs becomes large, the influence of every individual SBS can be modeled with the effect of the collective (aggregate) behavior of the SBSs. 
Mean-field game (MFG) theory enables us to transform these multiple interactions into a single aggregate interaction, called MF interaction, via MF approximation. 
According to \cite{MF_ref2}, this approximation holds under the following conditions: 
(i) a large number of players, (ii) exchangeability of players under the caching control strategy and (iii) finite MF interaction.

Remark that the first condition corresponds to the definition of UDNs.
Players in the SDG are said to be exchangeable or indistinguishable under the control ${p}_{k,j}(t)$ and the states of players and contents, 
if the player's control is invariant by their indices and decided by only their individual states.
In other words, permuting players' indices cannot change their control policies. Under this exchangeability, we can focus on a generic SBS by dropping its index~$k$. 

The MF interactions \eqref{interaction_1} and \eqref{interaction_2} should asymptotically converge to a finite value under the above conditions.
MF overlap \eqref{interaction_1} goes to zero when the number of contents per SBS is extremely large, i.e. $M \gg N $. Such a condition implies that the cardinality of the set consisting of asymptotically equal content popularity goes to infinity. In other words, $N_{r(j)}$ goes to infinity yielding that the expected amount of overlapping content per unit storage size $I^r_{k,j}(\boldsymbol{p}_{-k,j}(t))$ becomes zero. 
In terms of interference, MF interference converges  as
the ratio of SBS density to user density goes to infinity, i.e.  $N_a\lambda_b^{\alpha}/(\lambda_u R)^4 \rightarrow \infty$ \cite{JHP1}. Such a condition corresponds to the notion of UDN  \cite{interf_udn} or massive MIMO ($N_a \rightarrow \infty$). That is, the conditions enabling MF approximation  to inherently hold under ultra-dense caching networks.

To approximate the interactions from other SBSs, we need a state distribution of SBSs and contents at time $t$, called MF distribution $m_t(x(t),Q(t))$.
This is defined by a counting measure $M_t^{(N\times M)}(x(t),Q(t))= \frac{1}{NM}\sum_{j=1}^{M}\sum_{k=1}^{N} \delta_{\{x_j(t),Q_k(t)\}}$.
When the number of SBSs increases, the empirical distribution $M_t^{(N \times M)}(x_j(t),Q(t))$ converges to $m_t(x_j(t),Q(t))$, which is the density of contents and SBSs in state $(x(t),Q(t))$. Note that we omit the SE $\mathcal{R}(t)$ from the density measure to only consider temporally correlated state without loss of generality.

To this end, we utilize a Fokker-Planck-Kolmogorov (FPK) equation \cite{MFG_application} that is a partial differential equation capturing the time evolution of the MF distribution $m_t(x_j(t),Q(t))$ under dynamics of the popularity $x_j(t)$ and the remaining storage size $Q(t)$. 
The FPK equation for $m_t(x_j(t),Q(t))$ subject to the temporal dynamics  \eqref{SDE_x} and  \eqref{SDE_q} are given as follows:
\begin{align}
&0= \partial_t m_t(x_j(t),Q(t)) +r(\mu_j-x_j(t))\partial_x m_t(x_j(t),Q(t)) \nonumber \\&+ (e_j\!-\!L_j p_{j}(t)) \partial_{Q}m_t(x_j(t),Q(t)) 
 \!- \!\frac{\eta^2}{2}\partial_{xx}^2 m_t(x_j(t),Q(t)).  \label{fpk_1}
\end{align}
\noindent Let us denote the solution of the FPK equation \eqref{fpk_1} as $m_t^*(x_j(t),Q(t))$. 
Exchangeability and existence of the MF distribution allow us to approximate the interaction $I^r_{k,j}(\boldsymbol{p}_{-k,j}(t))$ as a function of $m_t^*(x_j(t),Q(t))$ as follows:
\begin{equation}
I^r_{j}(t,m_t^*(x_j(t),Q(t)\!)\!)\!=\!\!\int_{\!Q}\!\int_{\!x} \frac{m_t^*(x_j,Q){p}_{j}(t,\!x(t),\!Q(t))}{C_{k,j}N_{r(j)}} \text{d}x\text{d}Q, \label{MF_interaction_apprx}
\end{equation}
\noindent Remark that we can estimate the interaction from \eqref{MF_interaction_apprx} without observing other SBSs' caching strategies.
Thus, it is not necessary to have full knowledge of the states or the caching control policies of other SBSs. It means that an SBS only needs to solve a pair of equations, namely the FPK equation \eqref{fpk_1} and the following modified HJB one obtained by applying the MF approximation \eqref{MF_interaction_apprx} to \eqref{hjb_sdg}:  
\begin{align}
0&\!= \partial_t v_{j}(t)\! +\!\!\mathop{\text{inf}}\limits_{p_{j}(t)} \!\bigg{[}\!J_{j}(p_{j}(t),I_j(t,m_t^*(x_j(t),Q(t)\!)\!)\!+\! \frac{\eta^2}{2}\partial_{\small{xx}}^2 v_{j}(t)\nonumber \\
 &+ (e_j -L_j p_{j}(t)) \partial_{Q}v_{j}(t)+r(\mu-x_j(t))\partial_x v_{j}(t)\bigg{]}. \label{hjb_mfg}
\end{align}

The optimal trajectory of the LRA cost  $v^*_{j}(t)$ is found by applying backward induction to the single HJB equation \eqref{hjb_mfg}. Also, its corresponding MF distribution (state distribution) $m_t^*(x_j(t),Q(t))$ is obtained by forward solving the FPK equation \eqref{fpk_1}. These two solutions  $[m_t^*(x_j(t),Q(t)), v_j^*(t)]$ define the mean-field equilibrium (MFE), defined as follows:

\vskip 5pt
\noindent {\bf Definition 2}: The generic caching strategies $p_{j}^*(t)$ achieves a  MFE, if for all adimissible caching strategy set $p_{j}(t)$, it is satisfied that 
\begin{align}
\mathcal{J}_{j}(p_{j}^*(t), m^*_t(x_j(t),Q(t))\leq \mathcal{J}_{j}(p_{j}(t), m^*_t(x_j(t),Q(t)), \label{Def_NE}
\end{align}
under the temporal dynamics \eqref{const_1} and \eqref{const_2} for a initial MF distribution $m_0$.
The MFE corresponds to the $\epsilon$-Nash equilibrium:
\begin{align}
\mathcal{J}_{k,j}(p_{k,j}^*(t), \boldsymbol{p}_{-k,j}^*(t))\leq \mathcal{J}_{j}(p_{j}^*(t), m^*_t(x_j(t),Q(t)) -\epsilon, 
\end{align}
where $\epsilon$ asymptotically becomes to zero for a sufficiently large number of SBSs.

Let us define $p_j^*(t)$ as an optimal caching control strategy which achieves the MFE yielded by the optimal caching cost trajectory $v_j^*(t)$ and MF distribution $m_t^*(x_j(t),Q(t))$.
The optimal solution $p_j^*(t)$ is given by the following Proposition~1.
%
%and find the solution of the stochastic optimization problem \textbf{P1}.  

\vskip 10pt
 \noindent {\bf Proposition 1.} {\it The optimal caching amount is given by: }
\begin{eqnarray}
p_{j}^*(t)=\frac{1}{L_j}\left[B_{j}(t)- \frac{1+I^r_j(t,m^*_t(x_j(t),Q(t)))}{\mathcal{R}(t,I^f(t)) x_j(t)\partial{\scriptscriptstyle 
 {Q}}{v^*_{j}}}    \right]^+,  \label{Propo1}
 \end{eqnarray}
 {\it where $m_t^*(x(t),Q(t))$ and $v_j^*(t)$ are the unique solutions of  \eqref{fpk_1} and \eqref{hjb_mfg}, respectively. }  
 \vskip 10pt
\noindent {\it Proof}: The optimal control control of the differential game with HJB equations  is the argument of the infimum term \eqref{hjb_mfg}~\cite{exist_HJBsol1}.
\begin{align}
p_j^*(t)&\!=\!\arginf\limits_{p_{j}(t)} \bigg{[}J_{j}(p_{j}(t),\!I_j(t,m_t^*(x_j(t),Q(t)))\!+\! \frac{\eta^2}{2}\partial_{xx}^2 v_{j}(t)\nonumber \\
 &+\! (e_j -L_j p_{j}(t)) \partial_{Q}v_{j}(t)+r(\mu-x_j(t))\partial_x v_{j}(t)\bigg{]} \label{infimum}
\end{align}
The infimum term \eqref{infimum} is a convex function of $p_j(t)$ for all time $t$, since its first and second order derivative are lower than zero. Hence, we can apply Karush-Khun-Tucker (KKT) conditions and get a sufficient condition for the unique optimal control $p_j^*(t)$ by finding a critical point given by:
\begin{align}
\frac{\partial}{\partial p_j(t)} \big[J_{j}(p_{j}(t),I_j(t,m_t^*(x_j(t),Q(t))) \nonumber\\
+ (e_j -L_j p_{j}(t)) \partial_{Q}v_{j}(t) \big]= 0.\label{pppp}
\end{align}
Due to the convexity, the solution of equation \eqref{pppp} is the unique optimal solution described as follows:
\begin{equation}
p_{j}^*(t)=\frac{1}{L_j}\left[B_{j}(t)- \frac{1+I^r_j(t,m^*_t(x_j(t),Q(t)))}{\mathcal{R}(t,I^f(t)) x_j(t)\partial{\scriptscriptstyle 
 {Q}}{v^*_{j}}}    \right]^+
\end{equation}
Remark that $p^*_j(t)$ is a function of $m^*_t(x_j(t),Q(t))$ and $v^*_{j}$, which are solutions of the equations \eqref{fpk_1} and \eqref{hjb_mfg}, respectively.
The expression of $p_j^*(t)$ \eqref{pppp} provides the final versions of the HJB and FPK equations as follows:
{\small
\begin{align}
0&= \partial_t v_{j}(t) -\frac{\log \left(B_{j}(t)-\left[B_{j}(t)- \frac{1+I^r_j(t,m_t^*(x_j(t),\!Q(t)))}{\mathcal{R}(t,I^f(t))x_j(t)\partial{\scriptscriptstyle{Q}}{v_{j}}}    \right]^{\scriptscriptstyle{+}}\right)}{\mathcal{R}(t,I^f(t))x_j(t)} \nonumber \\
 &\!\times \!(1\!+\!I^r_j(t,\!m_t^*(x_j(t),\!Q(t)\!)\!)\!)+\!\frac{\alpha(C\!-\!Q(t))}{C}\!+\!r(\mu_j\!-\!x_j(t)\!)\partial_x v_{j}(t)\nonumber \\
 &\!+\! \left(\!e_j \!-\!\left[\!B_{j}(t)\!- \!\frac{\!1\!+\!I^r_j(t,m_t^*(x,\!Q))}{\mathcal{R}(t,I^f(t))x_j(t)\partial{\scriptscriptstyle{Q}}{v_{j}}}\!\right]^{\!\scriptscriptstyle{+}}\right) \!\partial_{Q}v_{j}(t)\!+\! \frac{\eta^2}{2}\partial_{xx}^2 v_{j}(t), \nonumber\\
 \nonumber\\
0&= \partial_t m_t(x_j(t),Q(t)) +r(\mu_j-x_j(t))\partial_x m_t(x_j(t),Q(t)) 
\nonumber\\&- \frac{\eta^2}{2}\partial_{xx}^2 m_t(x_j(t),Q(t))
\nonumber \\&+\! \left(\!e_j \!-\!\left[\!B_{j}(t)\!- \!\frac{\!1\!+\!I^r_j(t,m_t(x_j(t),Q(t)))}{\mathcal{R}(t,I^f(t))x_j(t)\partial{\scriptscriptstyle{Q}}{v_{j}^*}}\!\right]^{\!\scriptscriptstyle{+}}\right) \!\partial_{Q}m_t(x_j(t),\!Q(t)). \nonumber% \label{fpk_final}
\end{align}}

\noindent From these equations, we can find the values of
$v_j^*(t)$ and $m_t^*(x(t),Q(t))$. Note that the smoothness of the drift functions and in the dynamic equation and the cost function \eqref{inst_global_cost} assures the uniqueness of the solution \cite{exist_HJBsol1}. \hfill$\blacksquare$ 
\vskip 5pt

\begin{figure}
\centering
\includegraphics[angle=0, width=9cm]{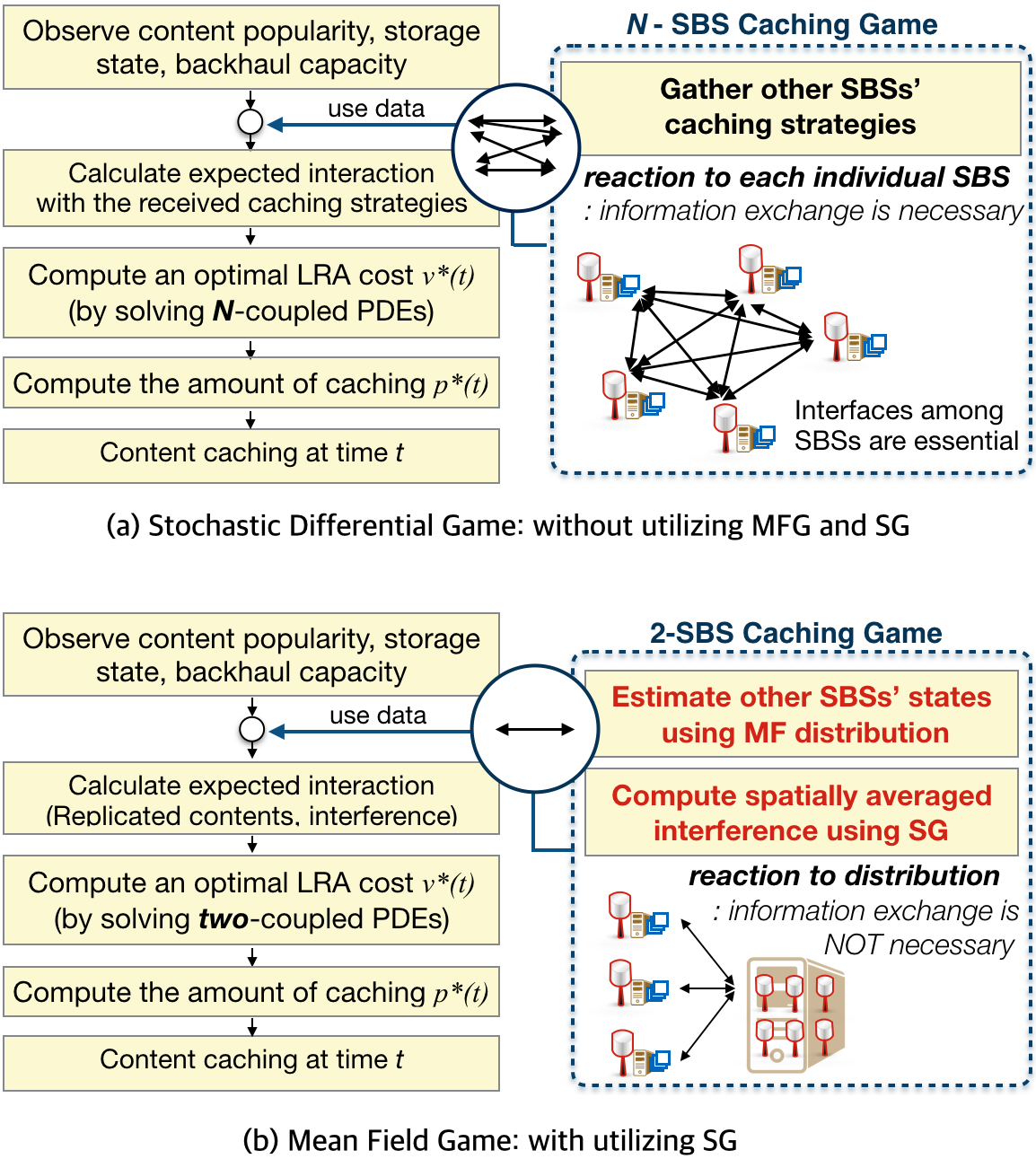}   %height=0.56
\caption{\small{Ultra-dense edge caching flow charts according to the approaches of SDG and MFG, respectively. (a) In the framework of SDG, we solve the game of $N$ SBSs (players)  reacting to each individual SBS. (b) By incorporating  MFG theory and SG into the framework, we can estimate the collective reaction of other SBSs. This relaxes the $N$-SBS caching game to a two-SBS caching game. }  }\label{diagram} 
\end{figure}

We obtain the optimal caching amount $p_{j}^*(t)$ in a water-filling fashion, where the backhaul capacity $B_j(t)$ determines the water level. Noting that the average rate per unit bandwidth $\mathcal{R}(t)$ increases with the number of antenna $N_a$ and SBS density $\lambda_b$, SBSs cache more contents from server when they can deliver content to users with high wireless capacity. 
Also, SBSs diminish the caching amount of content $j$, when the estimated amount of content overlap $I^r_j(t,m^*_t(x_j(t),Q(t)))$ is large.
%The partial derivative $\partial{\scriptscriptstyle{Q}}{v_{j}}$ 

Note that the existence and uniqueness of the optimal caching control strategy is guaranteed. The optimal caching algorithm converges to a unique MFE, when the initial conditions $m_0$, $x_j(0)$, and $Q(0)$ are given.
 The specific procedure of this MF caching algorithm is described in the following Algorithm~1.

\begin{algorithm}
\centering
\caption{Mean-Field Caching Control}\label{euclid}
\begin{algorithmic}[1]
\REQUIRE $x_j(t)$, $m_0$, $B(t)$ and $Q(0)$
\STATE  Find the optimal trajectory of caching cost and state distribution $[v_j^*(t),m_t^*(x_j(t),Q(t))]$ by solving HJB \eqref{hjb_mfg} and FPK \eqref{fpk_1} equations \vskip 3pt
\STATE Calculate  $I_j^r(t,m_t^*(x_j(t),\!Q(t)))$, $I^f(t)$ and $\partial{\scriptscriptstyle{Q}}{v^*_{j}}$
 \vskip 3pt
\STATE Compute the instantaneous caching amount $p_{j}^*(t)$\\ :$\quad p_{j}^*(t)=\frac{1}{L_j}\left[B_{j}(t)- \frac{1+I^r_j(t,m^*_t(x_j(t),Q(t)))}{\mathcal{R}(t,I^f(t)) x_j(t)\partial{\scriptscriptstyle 
 {Q}}{v^*_{j}}}    \right]^+$
 \vskip 3pt
\STATE Get values of $[x_j(t),Q(t)]$  according to the dynamics 
\STATE Go line 2
\end{algorithmic}
\end{algorithm}

The respective processes of solving \textbf{P1} in ways of SDG and MFG are depicted in Fig. \ref{diagram}.  
It should be remarked that the solution of the MFG becomes equivalent to that of the $N$-player SDG \textbf{P1} for large $N$.  
The complexity of the proposed method is much lower compared to solving the original $N$-player SDG  \textbf{P1}. It is worth mentioning that the number of PDEs to solve for one content is reduced to two from the number of SBSs $N$. It leads to maintaining the consistent complexity even though the number of player $N$ becomes large. 

\begin{figure}\centering
\includegraphics[angle=0, width=8.5cm]{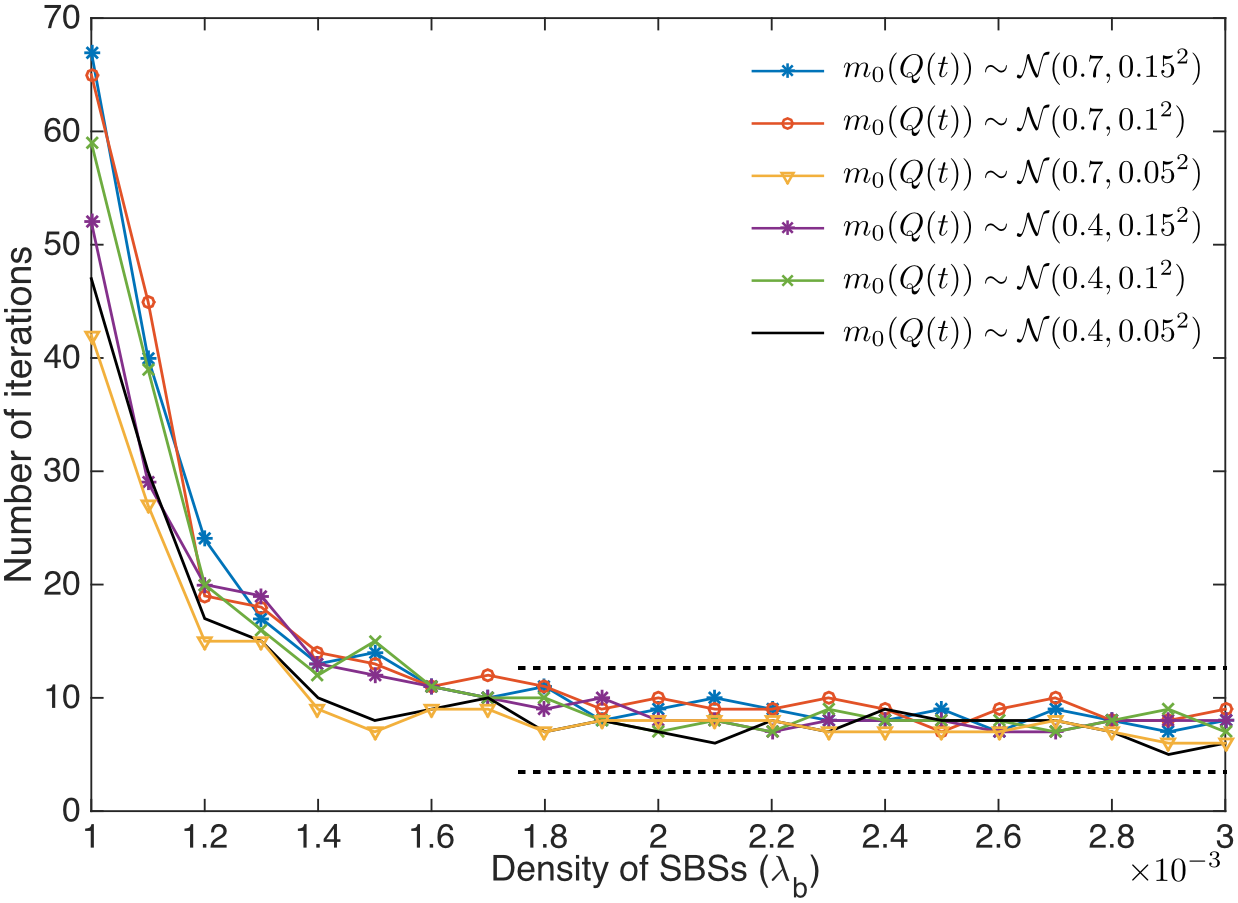}   %height=0.56
\caption{\small{The number of iterations that are required to solve the coupled HJB and FPK equations with respect to different densities of SBSs. Each point is a result for the different density of SBSs that are randomly located in a fixed area.   }  }\label{complexity_it}
\end{figure}

We verify this consistency of the computational complexity via simulations as shown in Fig. \ref{complexity_it}, which represents the number of iterations required to solve the HJB-FPK equations \eqref{hjb_mfg} and \eqref{fpk_1} as a function for different SBS densities $\lambda_b$. 
Here, we observe that for highly dense networks, the caching problem \textbf{P1} is numerically solved by within a few iterations. It means that the computational complexity remains consistent regardless of the SBS density $\lambda_b$, or the number of players $N$. This consistency of complexity holds for various initial storage state distribution of SBSs. It is worth mentioning that the number of iterations to reach the optimal caching strategy is bounded within tens of iterations even for low SBS density.

\section{Numerical Results}

In this section, we present numerical results  to evaluate the proposed algorithm  under spatio-temporal content popularity and network dynamics as illustrated in Fig.~\ref{system_model_CRP}. 
Let us assume that the initial distribution of the SBSs $m_0$ is given as a normal distribution. We assume that the storage size $Q(t)$ belongs to a set $[0,1]$ for all time $t$. Assuming Rayleigh fading with mean one, we set the parameters as shown in Table I.
In order to
solve the coupled PDEs (the first step of the Algorithm 1) using a finite element method, we used the MATLAB PDE solver.

%follows: 
%$\gamma=0.01,\lambda_u=0.0001,\lambda_b=0.003,R=10/\!\sqrt{\pi}, B(t)=1,N_{r(j)}=20, Q(0)=0.7, \eta=0.1,\alpha=4$.
%

\begin{table} [h]
\centering \caption{Key simulation parameters}\small
\small\begin{tabular}{|l||l|}
  \hline
  \footnotesize Parameter & \footnotesize Value  \\
  \hline
\footnotesize SBS density $\lambda_b$& \footnotesize 0.005, 0.02, 0.035, 0.05 (SBSs/km$^2$)
 \\
\footnotesize  User density $\lambda_u$  &\footnotesize $10^{-4}$, $2.5 \times 10^{-4}$ (users/km$^2$)
   \\
\footnotesize   Transmit power $P$ &\footnotesize 23 dBm
 \\
 \footnotesize  Noise floor &\footnotesize -70 dBm
 \\
\footnotesize   Number of contents    &\footnotesize 20
 \\
   CRP parameters  $\theta, \nu$  &\footnotesize $\theta=1,\nu=0.5$
   \\
\footnotesize  Reception ball radius $R$ &\footnotesize $10/\sqrt{\pi} $ km
 \\
\footnotesize  Network size &\footnotesize 20  km $\times$ 20 km
 \\
\footnotesize   File discarding rate $e_j$&\footnotesize 0.1
 \\
  \hline
\end{tabular}\label{table}
\end{table}

\subsection {Mean-field equilibrium achieved by the proposed MF caching algorithm}

\begin{figure} 
\centering
\includegraphics[angle=0, width=8.2cm]{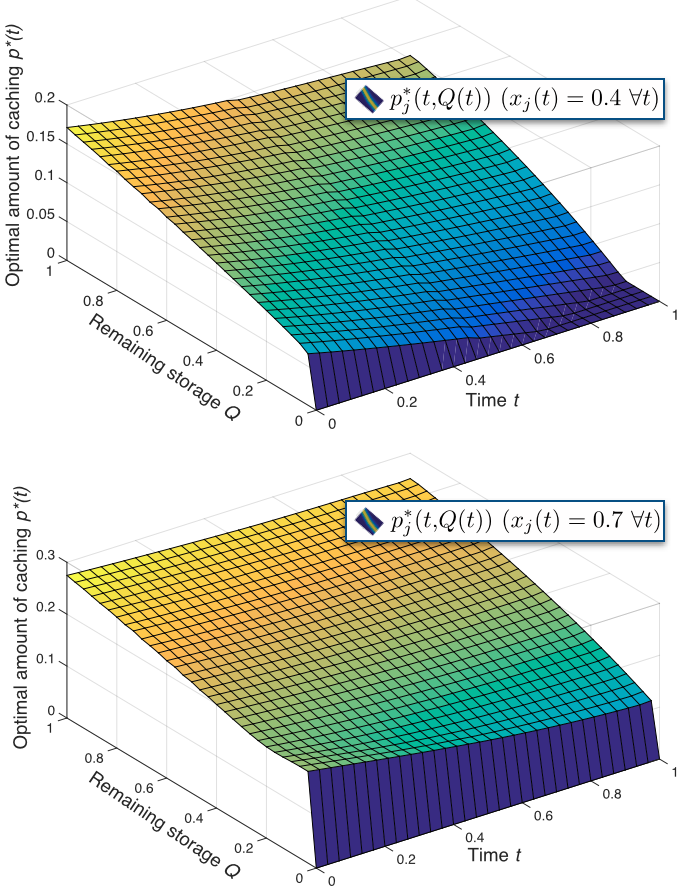}   %height=0.56
\caption{\small{Evolution of the optimal caching amount of a content $p^*(t)$ at the MF equilibrium under two different content popularities 0.4 and 0.7, assuming that the content popularity is static. The initial MF distribution $m_0(Q(0))$ is given as $\mathcal{N}(0.7,0.05^2)$.   }  }\label{trajectory_control}
\end{figure}

To demonstrate that the proposed MF caching algorithm achieves the MFE, we 
 assume full knowledge of contents request probability, implying that SBSs can get perfect popularity information.
We numerically analyze the algorithm performance when the content request probability is static in order to observe the trajectory of the caching algorithm and MF distribution. 
In this case, the problem becomes simplified, because the caching control strategies does not depend on the evolution law of the content popularity. 
Specifically, in HJB \eqref{hjb_mfg}  and FPK \eqref{fpk_1} equations, the derivative terms with respect to the content request probability $x$ become zero.

%\begin{figure}
%\centering
%\includegraphics[angle=0, height=0.33\textwidth]{LRAcost_0_3}   %height=0.33
%\caption{{\small Long run average costs of different caching strategies ($Q(0)=0.7, x(0) =0.3, \eta=0.1, a=0.15,u=0.1$).} }\label{LRA} 
%\end{figure}

Fig. \ref{trajectory_control} shows the evolution of the optimal caching amount $p^*(t)$ with respect to the storage state and time. We observe that the value of $p^*(t)$ is lower than the content request probability at all time slots for both different content popularities. It is for regulating the content overlap to prevent redundant backhaul and storage usage. Also, it is because SBSs save storage capacity for downloading more popular contents.

\begin{figure*}[ht]
\centering
\includegraphics[angle=0, width=0.99\textwidth]{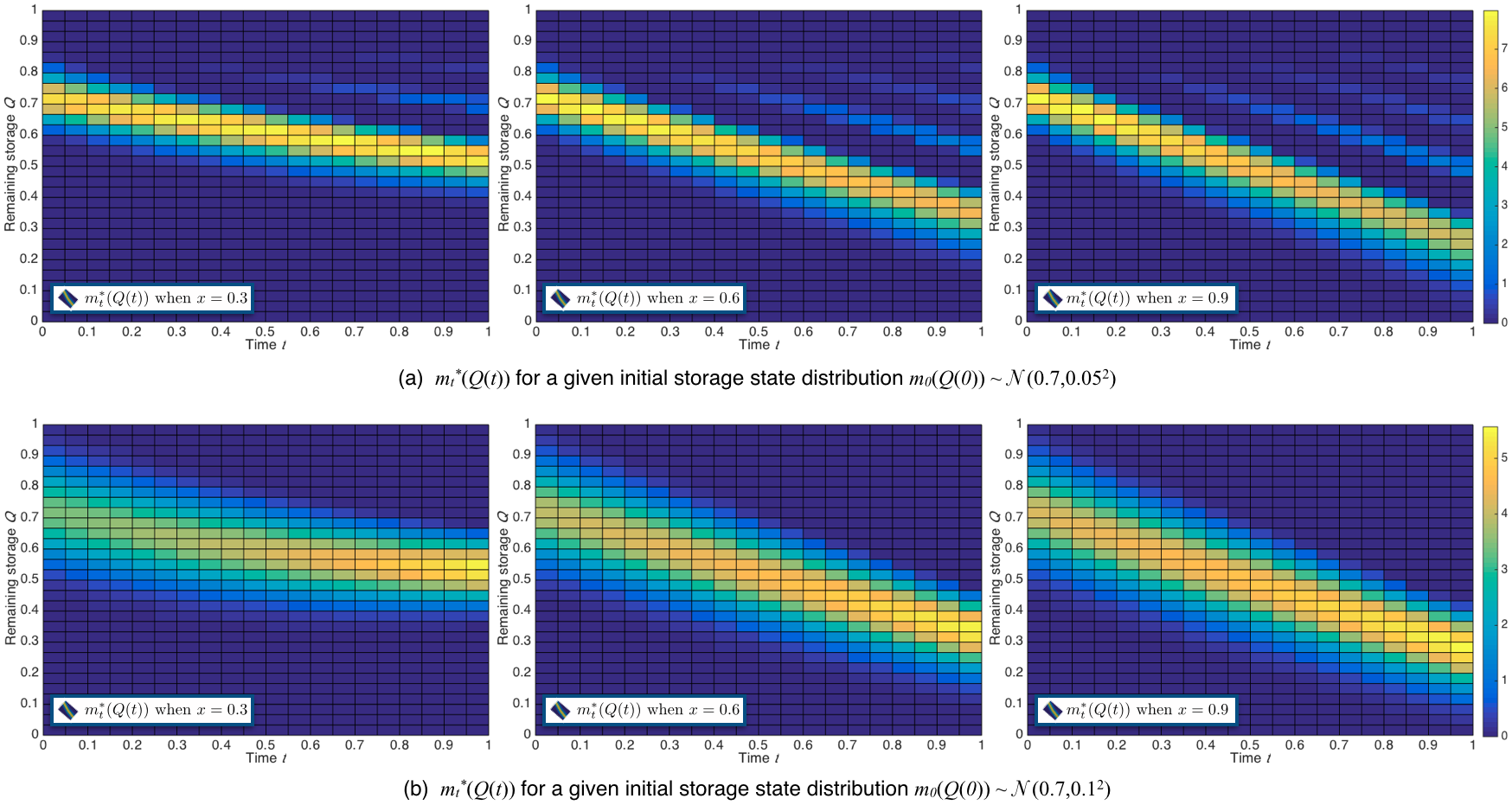} 
\caption{{\small A heat map illustration of the MF distribution $m_t^*(Q(t))$ that represents the instantaneous density of SBSs having the remaining storage space $Q(t)$  for an arbitrary content during a long-term period $\{\ 0\leq t \leq T\}$, when the proposed MF caching algorithm is applied. 
A bright-colored point means there are many SBSs with the unoccupied storage size corresponding to the point.
It shows the temporal evolution of the density of SBSs  with respect to different content popularity $x_j$, and initial distribution $m_0(Q(0))$ $(B(t)=1,N_{r(j)}=20, \lambda_u=0.001,\lambda_b=0.03)$.}}\label{MF_dist_total} 
\end{figure*}

Fig. \ref{MF_dist_total}  represents the instantaneous density of SBSs having the remaining storage size $Q(t)$ in terms of the MF distribution $m_t^*(Q(t))$ for a content during a period $\{0\leq t \leq T\}$, where $T=1$. A bright-colored point means there are many SBSs with the unoccupied storage size corresponding to the point. We observe that the unoccupied storage space of SBSs does not diverge from each other. This validates the fact that the proposed algorithm achieves the MFE.
At this equilibrium, the amount of downloaded content file decreases when the content popularity $x$ becomes low. This tendency corresponds to the trajectory of the optimal caching probability in Fig. \ref{trajectory_control}.
Almost every SBS has downloaded the content over time, but not used its entire storage. The remaining storage saturates even though the content popularity is equal to $0.9$. 
This implies that SBSs reduce the  downloading amount of the popular content in consideration of the expected content overlap amount, which increases with the popularity.

%Note that $m_t^*(Q)$ represents the density of the SBSs whose unoccupied storage size is equal to $Q$ at time $t$.

%\begin{figure}
%\centering
%\includegraphics[angle=0, height=0.3\textwidth]{MFcon_trajectory_x0_4_20}   %height=0.33
%\caption{Evolution of the optimal caching control $p^*(t)$ at the MF equilibrium when the content popularity is static ($x=0.6, B=1, Q=1, \mu=0.1$). The initial MF distribution $m_0$ is given as $\mathcal{N}(0.7,0.05^2)$. }\label{trajectory_control}
%\end{figure}
%

%\vskip -5pt
%\begin{figure}\centering
%\subfigure[$x=0.4$]{
%\includegraphics[angle=0, height=0.28\textwidth]{MFcon_trajectory_x0_4_20}}   %height=0.33
%\vskip -5pt
%\subfigure[$x=0.7$]{
%\includegraphics[angle=0, height=0.28\textwidth]{MFcon_trajectory_x0_7_20}}   %height=0.33
%\caption{Evolution of the optimal caching control $p^*(t)$ at the MF equilibrium when the content popularity is static ($x=0.6, B=1, Q=1, \mu=0.1$). The initial MF distribution $m_0$ is given as $\mathcal{N}(0.7,0.05^2)$. }\label{trajectory_control}
%\end{figure}

%\begin{figure}
%\centering
%\includegraphics[angle=0, height=0.3\textwidth]{LRA_0922}   %height=0.33
%\caption{Long run average costs of different caching strategies ($N=10, u=0.1, a=0.3, \mu=0.1, x(0) =0.3, \eta=0.1, B=1, Q(0)=0.3 $ ). }\label{LRA}
%\end{figure}

\subsection {Performance in terms of average caching cost}

\begin{figure}
\centering
\includegraphics[angle=0, width=8.7cm]{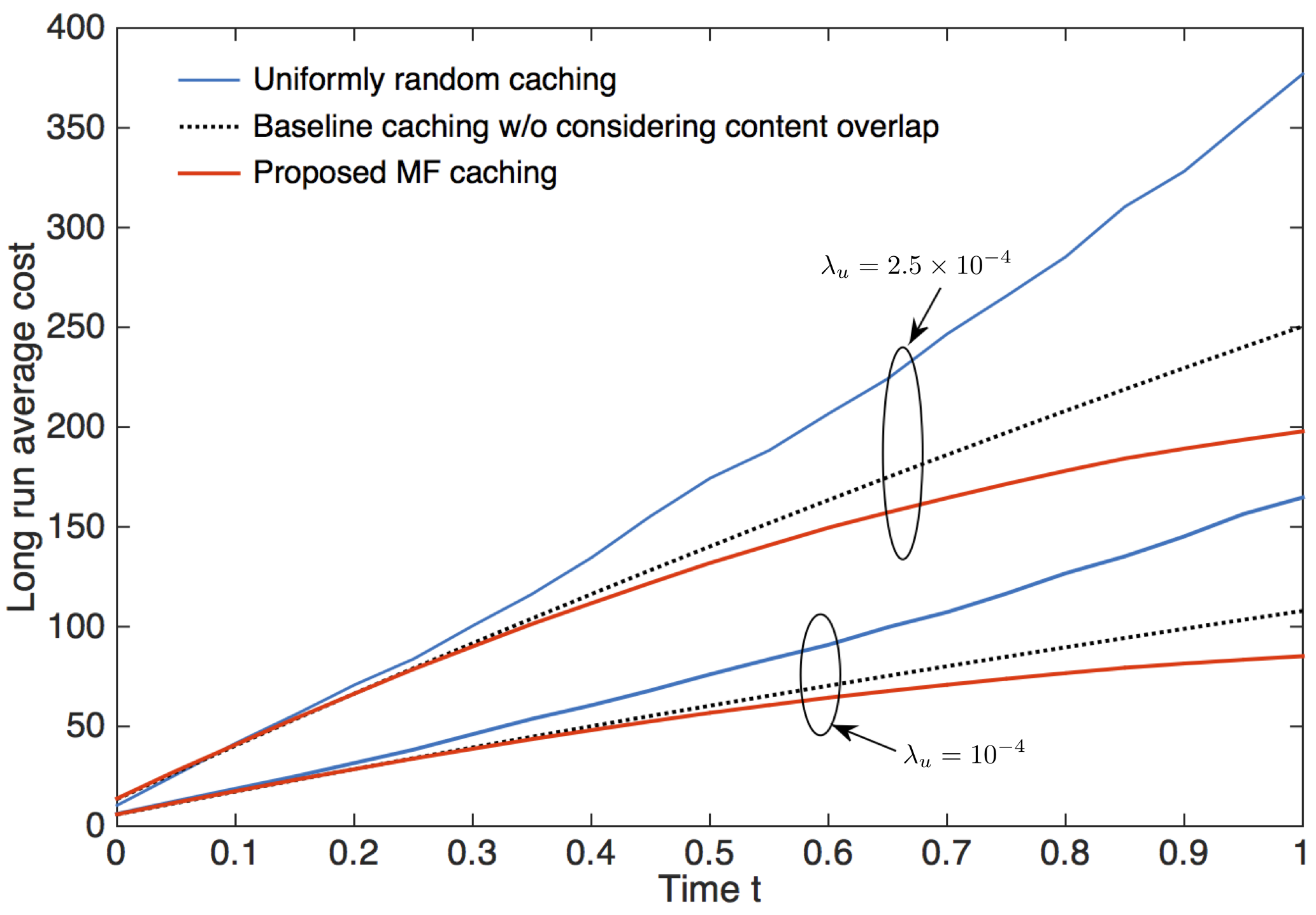}   %height=0.33
\caption{{\small Long run average costs of the caching strategies with respect to different user density $\lambda_u$. ($Q(0)=0.7, x(0) =0.3, \eta=0.1$).} }\label{LRA_user} 
\end{figure}

\begin{figure}
\centering
\includegraphics[angle=0, width=8.7cm]{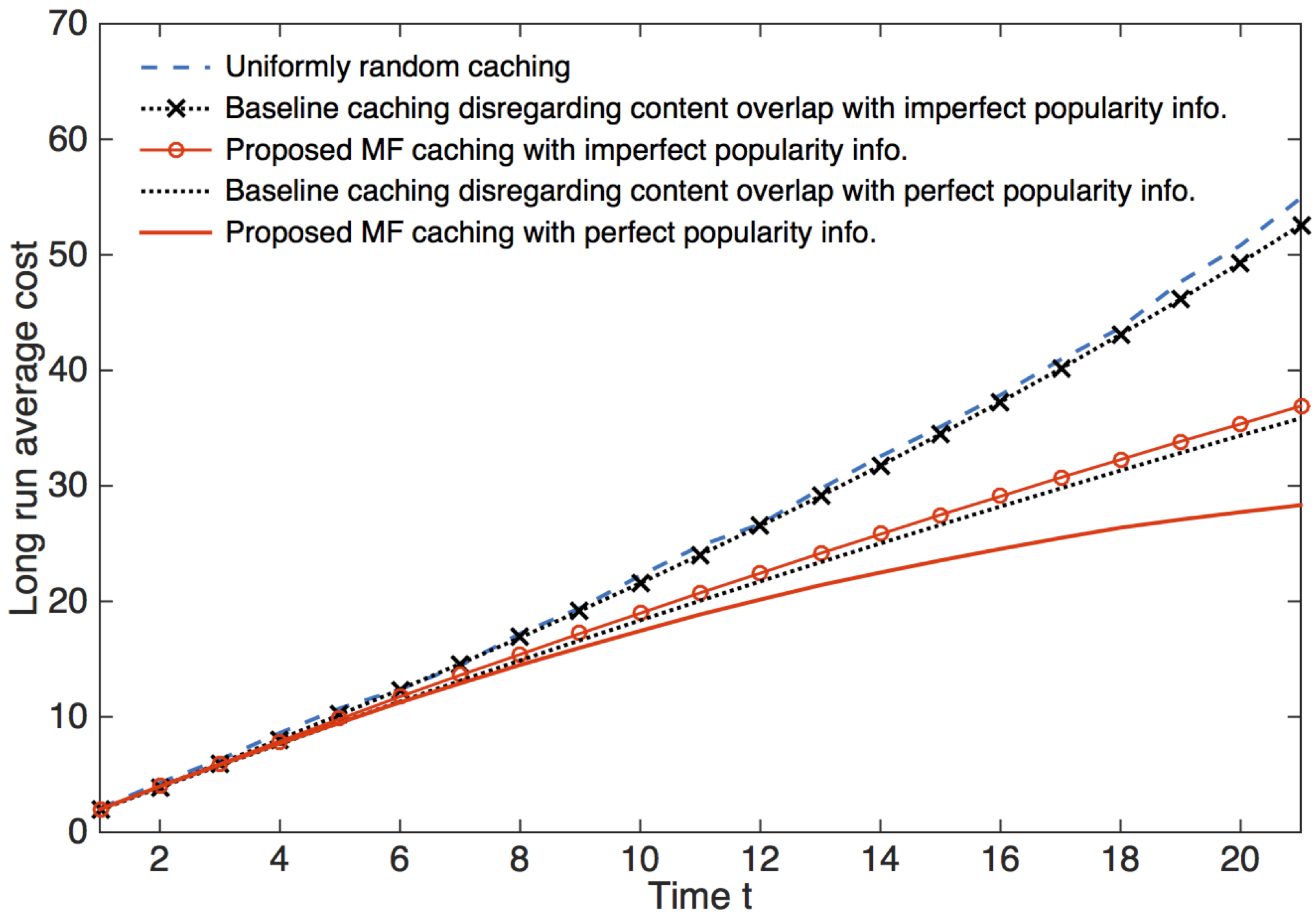}   %height=0.33
\caption{{\small Long run average costs of different caching strategies with perfect and imperfect popularity information. ($Q(0)=0.7, x(0) =0.3, \eta=0.1$).} }\label{LRA} 
\end{figure}

We investigate the performance of the proposed MF caching algorithm under spatio-temporal dynamics of the content popularity. 
Note that the temporal dynamics constrains the optimal caching control to follow the optimal trajectory $[v_j^*(t),m_t^*(x(t),Q(t))]$, a set of solutions to the coupled equations \eqref{hjb_mfg} and \eqref{fpk_1}. 
Additionally, we evaluate the robustness of our scheme to imperfect popularity information in terms of the LRA caching cost.
To this end, we compare the performance of the proposed MF caching algorithm with the following caching algorithms.
\begin{itemize}
\item {\it Baseline caching algorithm} that   does not consider the amount of content overlap but determines the instantaneous caching amount $\hat{p}_{j}(t)$ proportionally to the instantaneous request probability $x_j(t)$ subject to current backhaul, storage state, and interference described as follows: 
$\hat{p}_{j}(t)=\frac{1}{L_j}\left[B_{j}(t)- \frac{1}{1+\mathcal{R}(t,I^f(t)) x_j(t)   } \right]^+$. 
\item {\it Uniformly random caching} that determines the random caching amount following the uniform distribution.
\end{itemize}

In Fig.~\ref{LRA_user}, we compare the proposed MF caching algorithm, uniformly random caching, and the baseline caching  algorithm disregarding the content overlap among neighboring SBSs.
 The LRA costs over time for different user density $\lambda_u$ are numerically evaluated. The proposed caching control algorithm reduces about $24\%$ of the LRA cost as compared to the caching algorithm without considering the content overlap. This performance gain is due to avoiding the redundant content overlap and having an SBS under lower interference environment to cache more contents. As the user density $\lambda_u$ becomes higher for a fixed SBS density $\lambda_b$, the final values of the LRA cost increase for all the three caching schemes. When UDCNs are populated by numerous users, the fluctuation of  spatial dynamics of popularity increases and the number of  SBSs having associated users increase. 
Hence, both the aggregate interference imposed by the SBSs and the content popularity severely change over the spatial domain. In this environment,  the advantage of the proposed algorithm compared to the popularity based algorithm becomes larger, yielding higher gap between the final values of the produced LRA cost.

\begin{figure}
\centering
\includegraphics[angle=0, width=8.7cm]{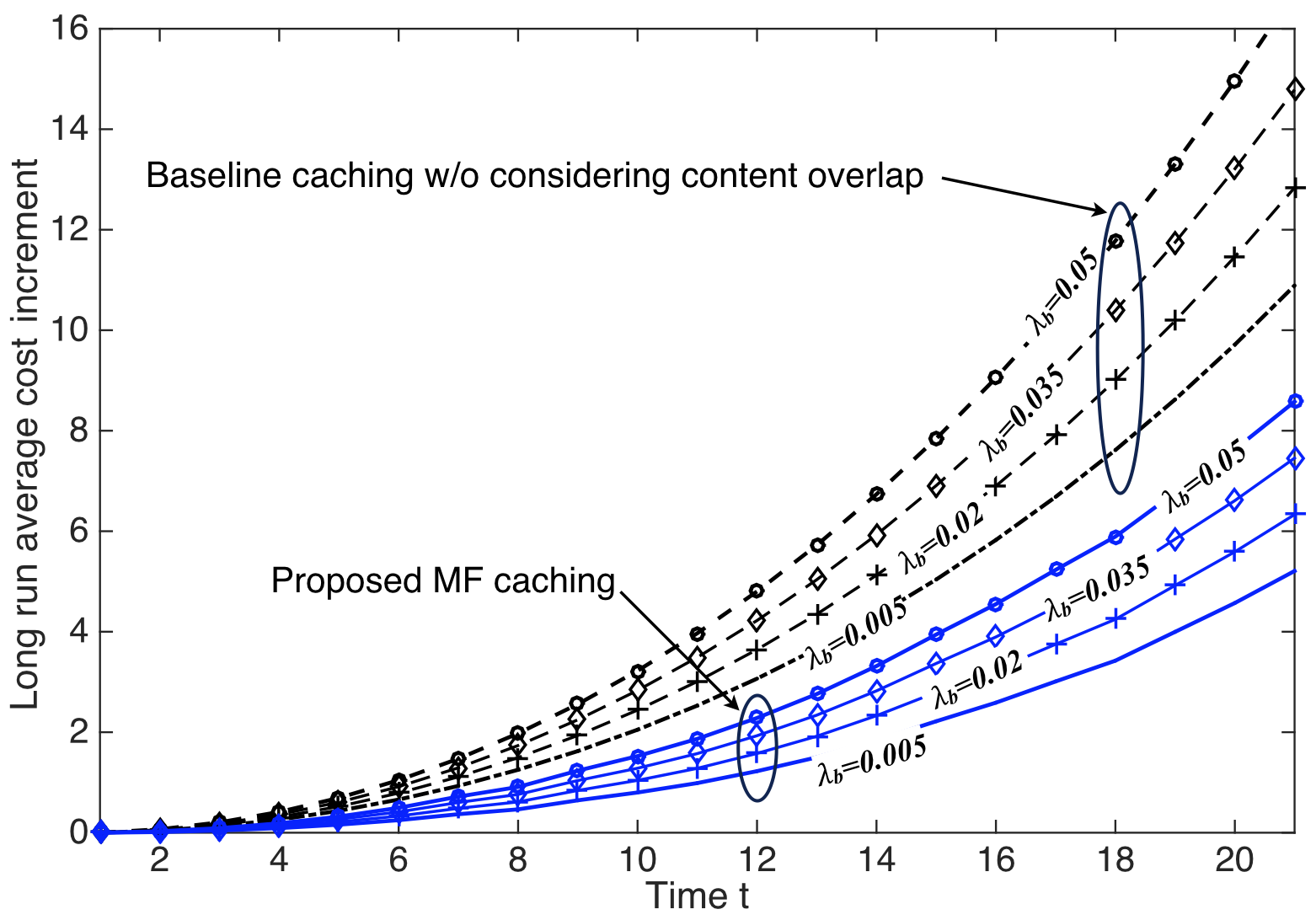}   %height=0.33
\caption{{\small Increment incurred by the imperfect popularity information in the long run average costs of the proposed MF caching and the caching without considering the content overlap with respect to different SBS density $\lambda_b$ ($Q(0)=0.7, x(0) =0.3, \eta=0.1$).} }\label{LRA_incre} 
\end{figure}

\begin{figure}
\centering
\includegraphics[angle=0, width=8.7cm]{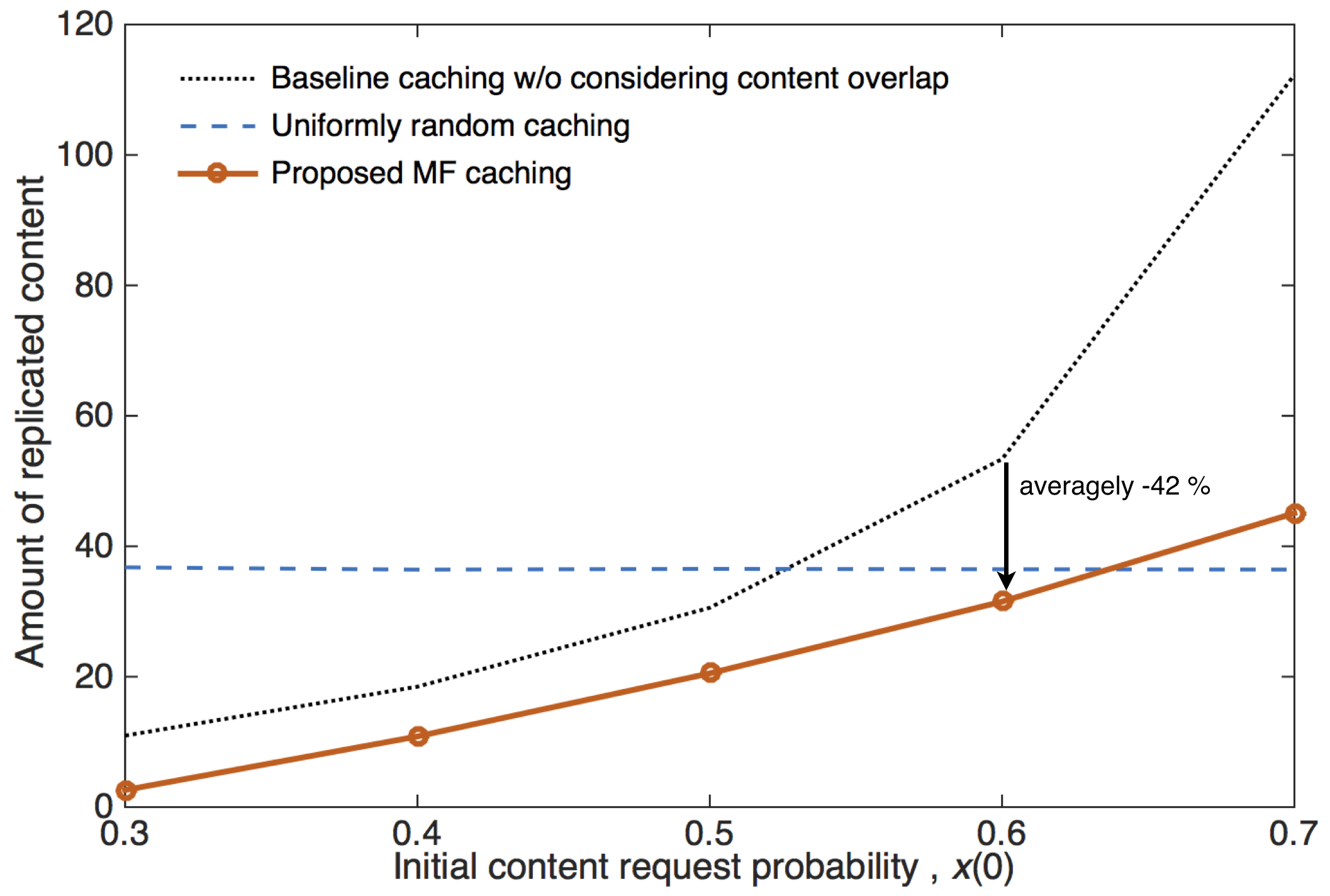}   %height=0.33
\caption{{\small The amount of overlapping contents per storage usage ($Q(0)=0.7, \eta=0.1$). } }\label{Repli}
\end{figure}

\textbf{Impact of prediction errors}: In real systems, perfect popularity information may not be available for SBSs due to misprediction or estimation error of content popularity. Hence, we assess how the proposed algorithm is robust to imperfect popularity information (IPI). For that, we model the IPI as follows:
\begin{equation}
\hat{x}(t)=x(t)+\Delta(t), \label{IPI_eq}
\end{equation}
where $\hat{x}(t)$ denotes a content request probability which is an SBS estimates, and $\Delta(t)$ denotes an instantaneous observation error for the request probability $x(t)$. It is worth noting that an SBS has perfect popularity information (PPI) if $\Delta(t)$ is equal to zero for all $t$ (i.e. $\hat{x}(t)=x(t)$). The magnitude of $\Delta$ determines accuracy of the popularity. 
 
Let us assume that an observation error $\Delta$ follows a normal distribution $\mathcal{N}(0.2,0.001^2)$. SBSs  determine their own caching control strategies based on imperfect content request probability $\hat{x}(t)$ \eqref{IPI_eq} instead of PPI $x(t)$. Under this IPI, we investigate the performance comparison in terms of the LRA caching cost over time as shown in Fig. \ref{LRA}. The impact of IPI increases with the number of SBSs because redundant caching occurs at several SBSs.
We compare the LRA increment of our MF caching algorithm with that of the popularity based algorithm for the different number of neighboring SBSs as shown in Fig. \ref{LRA_incre}. 
The numerical results demonstrate that the proposed algorithm is more robust to imperfect information of content popularity in comparison with the popularity based algorithm. Specifically, our caching strategy reduces about $50\%$ of the LRA cost increment as compared to the popularity based algorithm.

Fig. \ref{Repli} shows the amount of overlapping contents per storage usage as a function of the initial content probability $x(0)$. The proposed MF caching algorithm reduces caching content overlap averagely 42\% compared to popularity based caching. However, MF caching algorithm yields the higher amount of content overlap than random caching does when the content request probability becomes high. The reason is that the random policy downloads contents regardless of their own popularity, so the amount of content ovelap remains steady. On the other hand, MF caching increases the downloaded volume of popular contents.

\section{Concluding Remarks}

In this paper, we propose an edge caching algorithm for UDNs, taking into account the spatio-temporal user demand and network dynamics. For this, we devise a stochastic geometric network model and a spatio-temporal user demand model that specifies the content popularity changes within long-term and short-term duration.
We exploit an inherent feature of UDNs in that a large number of neighboring SBSs exists. 
When the number of SBSs becomes large, the influence of every individual SBS can be modeled with the effect of the aggregate behavior of the SBSs by leveraging
 the framework of MFG theory and SG. This enables SBSs to distributively determine their own caching control strategies without full knowledge of other SBSs' caching strategies or state. 
Hence, the proposed algorithm has low complexity that is independent of the SBS density. 
The proposed MF caching algorithm reduces not only the long run average cost but also the overlapping content amount compared to a caching strategy, which is merely based on content popularity. Furthermore, the MF caching algorithm is robust to imperfect information of content popularity in comparison with  a baseline caching algorithm without considering content overlap.

\ifCLASSOPTIONcaptionsoff
  \newpage
\fi

% that's all folks
\end{document}